\documentclass[%
 amsmath,
 amssymb,
 aps,
 pre,
 twocolumn
]{revtex4-1}

\usepackage{graphicx}
\usepackage{dcolumn}
\usepackage{bm}
\usepackage{appendix}

\usepackage{xr}
\begin{document}

\newcommand{\boldeta}{\eta} 
\newcommand{\boldlambda}{{\boldsymbol{\lambda}}}
\newcommand{\boldx}{{{x}}}
\newcommand{\trajx}{X}
\newcommand{\boldc}{c}

\title{Time-Asymmetric Fluctuation Theorem and Efficient Free Energy Estimation}
\author{Adrianne Zhong$^*$ and Benjamin Kuznets-Speck }
\email{adrizhong@berkeley.edu, biophysben@gmail.com. \\These authors contributed equally.}
\affiliation{Department of Physics, University of California, Berkeley, CA, 94720, USA \looseness=-1}
\affiliation{Biophysics Graduate Group, University of California, Berkeley, CA, 94720, USA \looseness=-1}
\author{Michael R. DeWeese}
\affiliation{%
 Department of Physics, University of California, Berkeley, Berkeley, CA, 94720
}%
\affiliation{%
Redwood Center For Theoretical Neuroscience and Helen Wills Neuroscience Institute, University of California, Berkeley, Berkeley, CA, 94720
}%

\date{\today}

\begin{abstract}
The free-energy difference $\Delta F$ between two high-dimensional systems is notoriously difficult to compute, but very important for many applications such as drug discovery ~\cite{cournia2017relative}. We demonstrate that an unconventional definition of work introduced by Vaikuntanathan and Jarzynski (2008) satisfies a microscopic fluctuation theorem that relates path ensembles that are driven by protocols unequal under time-reversal. It has been shown before that counterdiabatic protocols---those having additional forcing that enforces the system to remain in instantaneous equilibrium, also known as escorted dynamics or engineered swift equilibration---yield zero-variance work measurements for this definition. We show that this time-asymmetric microscopic fluctuation theorem can be exploited for efficient free energy estimation by developing a simple (i.e., neural-network free) and efficient adaptive time-asymmetric protocol optimization algorithm that yields $\Delta F$ estimates that are orders of magnitude lower in mean squared error than the generic linear interpolation protocol with which it is initialized.
\end{abstract}

\maketitle

\section{Introduction}

Free energy differences $\Delta F = F_B - F_A$ between pairs of potential energy functions $U_A(\boldx)$ and $U_B(\boldx)$ are sought after by physicists, chemists, and pharmaceutical scientists alike \cite{cournia2017relative, tawa1998calculation, kelly2006aqueous, chipot2007free, shirts2010free, cournia2021free}. Here, $\boldx \in \mathbb{R}^d$ is the configuration space coordinate, and the free energy for each potential is defined as $F_{A,B} = -\beta^{-1} \ln \int e^{-\beta U_{A,B}(\boldx)}\mathrm{d}\boldx$, where $\beta = (k_\mathrm{B} T)^{-1}$ is inverse temperature. For high dimensional systems, $\Delta F$ can only be calculated numerically through sampling methods, which can be computationally costly and slow to converge \cite{chipot2007free}. Here we present an adaptive method that greatly reduces the variance of $\Delta F$ estimates, based on a new fluctuation theorem we derive.

One class of estimators takes work measurements as input from protocols $U(\boldx, t)$ that ``switch'' $U(\boldx, 0) = U_A(\boldx) \rightarrow U(\boldx, t_f) = U_B(\boldx)$ in finite time $t_f$. Because the work, traditionally defined for a trajectory $\trajx(t)|_{t\in[0,t_f]}$ \cite{trajectorynotation} as  
\begin{equation}
W_\mathrm{trad}[\trajx(t)] = \int_0^{t_f} \frac{\partial U}{\partial t} (\trajx(t), t) \, \mathrm{d}t \label{eq:work-trad}
\end{equation}
satisfies the Jarzynski equality
\begin{equation}
    \langle e^{-\beta W} \rangle = e^{-\beta \Delta F}, \label{eq:jarzynski-equality}
\end{equation}
the Jarzynski estimator $\widehat{\Delta F}_\mathrm{Jar} = -\beta^{-1} \ln \, \big( n_s^{-1} \sum_{i=1}^{n_s} e^{-\beta W^i_\mathrm{trad}}\big)$ may be applied to work measurements $\{W^i_\mathrm{trad}, | i =1, .., n_s\}$. Unfortunately this estimator can be slow to converge, because the average is often dominated by rare events.

Estimators that use bi-directional work measurements (i.e., those that also consider $U_B \rightarrow U_A$ switching processes) generally have lower variance than uni-directional work estimators \cite{shirts2005comparison}. In particular, Shirts et al. in \cite{shirts2003equilibrium} showed that if forward  $\{W^i_F \, | \, i =1, .., n_s\}$ and reverse work measurements $\{W^i_R \, | \, i =1, .., n_s\}$, assumed here to be equal in number for simplicity, are collected from forward and reverse protocols satisfying Crooks Fluctuation Theorem\begin{equation}
    \mathcal{P}_F(+W) = \mathcal{P}_R(-W) \, e^{\beta (W - \Delta F)}, \label{eq:fluctuation-theorem}
\end{equation}
then the Bennett acceptance ratio estimator $\widehat{\Delta F}_\mathrm{BAR}$ \cite{bennett1976efficient}, defined implicitly as the $\Delta F$ satisfying
\begin{equation}
  \sum_{i = 1}^{n_s} \frac{1}{1 + e^{-\beta(W^i_F - \Delta F)}} -   \sum_{j = 1}^{n_s}  \frac{1}{1 + e^{-\beta(W^j_R + \Delta F)}} = 0, \label{eq:BAR-estimator}
\end{equation}
is the lowest-variance asymptotically-unbiased estimator. Bi-directional measurements of $W_\mathrm{trad}$ for a pair of time-reversal-symmetric forward and reverse protocols satisfy Eq.~\eqref{eq:fluctuation-theorem} \cite{crooks1998nonequilibrium, crooks1999entropy}, but measurements can also be collected from \textit{mixtures} of different measurement-protocol pairs
\begin{equation}
    \mathcal{P}_F(\cdot) = \sum_i \alpha_i \mathcal{P}^{i}_F(\cdot) \quad\mathrm{and}\quad \mathcal{P}_R(\cdot) = \sum_i \alpha_i \mathcal{P}^{i}_R(\cdot) \label{eq:mixtures}
\end{equation}
with $\sum_i \alpha_i = 1$, as long as each $(\mathcal{P}^{i}_F, \mathcal{P}^{i}_R)$ pair satisfies Eq.~\eqref{eq:fluctuation-theorem}.

In this paper, we consider the non-standard definition of work introduced in \cite{vaikuntanathan2008escorted}, for which there exists finite-time counterdiabatic driving protocols that give zero-variance work measurements. We explicitly show that it satisfies the fluctuation theorem Eq.~\eqref{eq:fluctuation-theorem} for measurements that are produced from forward and reverse protocols that are \textit{unequal} under time-reversal. We demonstrate that the time-asymmetric fluctuation theorem for this unconventional work may be exploited for efficient free energy estimation by proposing an algorithm that iteratively improves time-asymmetric protocols from bi-directional measurements collected across all iterations. On three examples of increasing complexity, we show that $10^3$ bi-directional measurements made under our adaptive protocol algorithm give $\Delta F$ estimates that are a factor of $\sim 10^2 - 10^4$ lower in mean squared error than the same number of bi-directional measurements made with the generic time-symmetric linear interpolation protocol with which it was initialized. 

The first version of this paper was posted on ArXiv in April 2023 \cite{zhong2023time}. Near-simultaneously, the preprint \cite{vargas2023transport} was posted on ArXiv, in which the authors independently derived the same theoretical results as we found for overdamped dynamics, and demonstrated through impressive numerical results the utility of the time-asymmetric fluctuation theorem.

\section{Time-asymmetric work}

For our setting we consider a time-varying potential energy $U_0(\boldx, t)$ for $t \in [0, t_f]$, that begins at $U_0(\boldx, 0) = U_A(\boldx)$ and ends at $U_0(\boldx, t_f) = U_B(\boldx)$. To this we add an additional potential $U_1(\boldx,t)$ that satisfies $U_1(\boldx, 0) = U_1(\boldx, t_f) = 0$. In the overdamped limit, a trajectory $\trajx(t)$ evolves according to the Langevin equation
\begin{equation} \label{eq:forward-langevin}
\dot \trajx(t) = -\nabla ( U_0 + U_1 ) + \sqrt{2\beta^{-1} } \,  \boldeta(t)\quad\mathrm{with}\quad \trajx(0) \sim \rho_A(\cdot).
\end{equation}
Here, $\rho_A(\boldx) = e^{-\beta [U_A(\boldx) - F_A]}$ is the equilibrium distribution for $U_A(\boldx)$, and $\boldeta(t)$ is an instantiation of standard $d$-dimensional Gaussian white noise specified by $\langle \eta_i(t) \rangle = 0$ and $\langle \eta_i(t)\eta_j(t') \rangle = \delta_{ij} \delta(t - t')$ \cite{gamma1}. (We consider underdamped dynamics in the SM \cite{SM}.)

In \cite{vaikuntanathan2008escorted} the authors introduced an unconventional work definition, which in our setting is the trajectory functional
\begin{equation} \label{eq:time-asymmetric-work} 
  W[\trajx(t)] = \int_{0}^{t_f} \frac{\partial U_0}{\partial t} - \nabla U_0 \cdot \nabla U_1 + \beta^{-1} \nabla^2 U_1 \, \mathrm{d}t 
\end{equation}
($\nabla^2$ is the scalar Laplace operator), and demonstrated that, remarkably, $W[\trajx(t)] = \Delta F$ for \textit{every} trajectory $\trajx(t)$, if $U_1(\boldx, t)$ gives the counterdiabatic force for $U_0(\boldx, t)$, meaning
\begin{equation}
  \frac{\partial \rho_0}{\partial t} = \nabla \cdot (\rho_0 \nabla U_1) \quad\mathrm{for}\quad \rho_0(\boldx, t) := e^{-\beta [U_0(\boldx, t) - F_0(t)]}. \label{eq:sufficient-condition}
\end{equation}
Here $\rho_0(\boldx, t)$ is the instantaneous equilibrium distribution corresponding to $U_0(\boldx, t)$, with time-dependent free energy $F_0(t) = -\beta^{-1} \ln \int e^{-\beta U_0(\boldx, t)} \mathrm{d}\boldx$ satisfying $F_0(0) = F_A$ and $F_0(t_f) = F_B$. Counterdiabatic driving has been studied before in various contexts \cite{del2013shortcuts, guery2019shortcuts, ilker2022shortcuts, iram2021controlling, frim2021engineered}. Under these conditions, the time-dependent probability distribution for Eq.~\eqref{eq:forward-langevin} is always in instantaneous equilibrium with $U_0(\boldx, t)$. 

Indeed, expanding Eq.~\eqref{eq:sufficient-condition} yields
\begin{equation}
  \frac{\partial U_0}{\partial t} - \nabla U_0 \cdot \nabla U_1 + \beta^{-1} \nabla^2 U_1 = \frac{\mathrm{d}F_0}{\mathrm{d}t}, \label{eq:integrand-reduction}
\end{equation}
which, when plugged into Eq.~\eqref{eq:time-asymmetric-work}, shows that the time-asymmetric work $W[\trajx(t)] = \int_0^{t_f} \dot{F_0}(t) \, \mathrm{d}t = F_0(t_f) - F_0(0) = \Delta F$ for \textit{every} trajectory $\trajx(t)$. With optimally chosen $U_0(\boldx, t)$ and $U_1(\boldx, t)$, the free energy difference may be obtained from simulating a single finite-time trajectory. Unfortunately, Eq.~\eqref{eq:integrand-reduction} is typically infeasible to solve for multidimensional systems, and to formulate the PDE, $\dot{F_0}(t)$, and therefore $\Delta F$, must already be known.

\section{Time-asymmetric microscopic fluctuation theorem}

In the late 1990s, Crooks \cite{crooks1998nonequilibrium, crooks1999entropy} discovered that the microscopic fluctuation theorem
\begin{equation}
  W[\trajx(t)] = \Delta F + \beta^{-1} \ln \frac{\mathcal{P}[\trajx(t)]}{\tilde{\mathcal{P}}[\tilde{\trajx}(t)]} \label{eq:work-MFT-definition}
\end{equation}
is satisfied by the traditional work $W = W_\mathrm{trad}$. Here $\mathcal{P}[\trajx(t)]$ is the probability of observing a trajectory $\trajx(t)$, and $\tilde{\mathcal{P}}[\tilde{\trajx}(t)]$ is the probability of observing its time-reversed trajectory $\tilde{\trajx}(t) = \trajx(t_f - t)$ in a ``reverse'' path ensemble driven by the protocol $\tilde{U}(\boldx, t) = U(\boldx, t_f - t)$. In this section, we derive the microscopic fluctuation theorem satisfied by the unconventional work definition Eq.~\eqref{eq:time-asymmetric-work}.

In our overdamped setting, the probability of realizing a trajectory $\trajx(t)$ from the dynamics Eq.~\eqref{eq:forward-langevin} may be formally expressed, up to a normalization factor, as 
\begin{equation}
\mathcal{P}[\trajx(t)] = \rho_A(\trajx(0)) e^{-\beta S[\trajx(t)]}, \label{eq:forward-path-probability}
\end{equation}
where
\begin{equation} 
    S[\trajx(t)] = (\mathrm{I}) \int_0^{t_f} \frac{|\dot{\trajx} + \nabla(U_0 + U_1)|^2}{4} \mathrm{d}t, \label{eq:forward-path-action}
\end{equation}
is the Onsager-Machlup action functional (see Appendix Sec.~\ref{appendix:OM-action} for a quick review, also \cite{adib2008stochastic}). We use  ($\mathrm{I}$) to indicate that the integral is taken in an Itô sense (reviewed in Appendix Sec.~\ref{appendix:stochastic-integrals}). After Eqs.~\eqref{eq:time-asymmetric-work} and \eqref{eq:forward-path-probability} are plugged into Eq.~\eqref{eq:work-MFT-definition}, straightforward manipulations under the rules of stochastic calculus (see Appendix Sec.~\ref{appendix:MFT-derivation}) yield

\begin{align}
    \tilde{\mathcal{P}}[\tilde{\trajx}(t)] &= e^{-\beta \{ U_A(\trajx(0)) - F_A + S[\trajx(t)] + W[\trajx(t)] - \Delta F\} } \nonumber \\  
    &= \rho_B(\tilde{\trajx}(0)) e^{-\beta \tilde{S}[\tilde{\trajx}(t)]  } , \label{eq:MFT}
\end{align}
where $\rho_B(\boldx) = e^{-\beta[U_B(\boldx) - F_B]}$ is the equilibrium distribution for $U_B(\boldx)$, and
\begin{equation}
\tilde{S}[\tilde{\trajx}(t)] =  (\mathrm{I}) \int_0^{t_f} \frac{|\dot{\tilde{\trajx}} + \nabla (\tilde{U}_0 - \tilde{U}_1)|^2}{4} \mathrm{d}t\label{eq:reverse-path-action} 
\end{equation}
has the form of a path action, with $\tilde{U}_{0,1}(\boldx , t) = U_{0,1}(\boldx , t_f - t)$ denoting the time-reversed potential energies. 
Eq.~\eqref{eq:MFT} gives the probability of observing the path $\tilde{\trajx}(t)$ under the Langevin equation
\begin{equation}
\dot{\tilde{\trajx}}(t) = -\nabla (\tilde{U}_0 - \tilde{U}_1) + \sqrt{2 \beta^{-1}} \boldeta(t) \quad\mathrm{with}\quad \tilde{\trajx}(0) \sim \rho_B(\cdot), \label{eq:reverse-langevin}
\end{equation}
which differs from Eq.~\eqref{eq:forward-langevin} by a minus sign on the $U_1$ term. In other words, the reverse path ensemble that satisfies Eq.~\eqref{eq:MFT} for the time-asymmetric work Eq.~\eqref{eq:time-asymmetric-work} is one that is driven by a protocol $\tilde{U}_0 - \tilde{U}_1$ that is \textit{different} from the time-reversal of the forward protocol $U_0 + U_1$. One can also verify that its associated definition of work 
\begin{equation}
  \tilde{W}[\tilde{\trajx}(t)] = \int_{0}^{t_f} \frac{\partial \tilde{U}_0}{\partial t} - \nabla \tilde{U}_0 \cdot \nabla (-\tilde{U}_1) + \beta^{-1} \nabla^2 (-\tilde{U}_1) \, \mathrm{d}t, \label{eq:reverse-work} 
\end{equation}
satisfies $\tilde{W}[\tilde{\trajx}(t)] = -W[\trajx(t)]$, so the same optimal $U_0(\boldx, t)$ and $U_1(\boldx, t)$ satisfying Eq.~\eqref{eq:sufficient-condition} also give $\tilde{W}[\tilde{\trajx}(t)] = -\Delta F$ for every trajectory. 

Through standard methods (see Appendix Sec.~\ref{appendix:MFT-FT}), the fluctuation theorem Eq.~\eqref{eq:fluctuation-theorem} follows directly from the microscopic fluctuation theorem Eq.~\eqref{eq:MFT}. Thus, the time-asymmetric work (Eq.~\eqref{eq:time-asymmetric-work}) holds a deeper significance than how it may first appear -- it relates the forward and reverse path ensembles given by Eqs.~\eqref{eq:forward-langevin} and \eqref{eq:reverse-langevin} that are driven by time-asymmetric protocols. 

Though much more involved, the time-asymmetric fluctuation theorem may also be derived for underdamped dynamics through similar techniques. We include our derivation in the SM \cite{SM}.

Significantly, the time-asymmetric fluctuation theorem may be exploited for efficient free energy estimation. In particular, by considering optimizing two \textit{different} protocols---one for the forward dynamics and the other for the reverse dynamics---the variance of $\Delta F$ estimates may be lowered by orders of magnitude. We now propose our algorithm demonstrating this.

\section{Algorithm}

In this section we present an on-the-fly adaptive importance-sampling protocol optimization algorithm, inspired by \cite{jie2010connection}, that uses the previously collected bi-directional samples to iteratively discover lower-variance time-asymmetric protocols. Exploiting the mathematical structure of the Onsager-Machlup action, our algorithm requires minimal computational overhead, solely the inclusion of easily-computable auxiliary variables in each trajectory's time-evolution.

Concretely, we consider the objective function 
\begin{align}
    J = J_F + J_R = \langle W \rangle_F + \langle \tilde{W} \rangle_R. \label{eq:objective-function} 
\end{align}
Jensen's inequality applied to Eq.~\eqref{eq:jarzynski-equality} implies $\langle W \rangle_F \geq \Delta F$ and $\langle \tilde{W} \rangle_R \geq -\Delta F$, with equality only for zero-variance optimal protocols. 

Our simulations are performed using the Euler-Mayurama method to discretize Eqs.~\eqref{eq:forward-langevin} and \eqref{eq:reverse-langevin}. Instead of directly discretizing Eq.~\eqref{eq:time-asymmetric-work}, we measure for every trajectory the expression derived from Eq.~\eqref{eq:MFT}

\begin{align}
    W[\trajx(t)] &= U_B(\trajx(t_f)) - U_A(\trajx(0)) + \beta^{-1} \ln \frac{\mathcal{P}[\trajx(t) | \trajx(0)]}{\tilde{\mathcal{P}}[\tilde{\trajx}(t) | \tilde{\trajx}(0)]} 
\end{align}
with the correct discrete-path probabilities, so as to preserve the fluctuation theorem \eqref{eq:work-MFT-definition}. In our setting this may be written as
\begin{equation}
    W[\trajx(t)] = \{ U_B(\trajx(t_f)) + \tilde{S}[\tilde{\trajx}(t)] \} - \{ U_A(\trajx(0)) + S[\trajx(t)] \}.   \label{eq:shadow-work}
\end{equation}

From now on we will use Einstein notation, where repeated upper and lower Greek indices signify summation. Let $\{U_\mu(\boldx, t) \, | \, \mu = 1, ..., M \}$ denote a set of time-dependent basis functions. Given the linear parameterization of the forward and reverse protocols $U_F = U_0 + U_1$ and $U_R = U_0 - U_1$
\begin{equation}
U_{F,R}(\boldx, t) = \begin{cases} U_A(\boldx) &\quad\mathrm{for}\quad t = 0 \\ 
\theta^\mu_{F,R} U_\mu(\boldx, t) &\quad\mathrm{for}\quad t \in (0, t_f) \\ 
U_B(\boldx) &\quad\mathrm{for}\quad t = t_f \\ 
\end{cases} \label{eq:linear-basis}
\end{equation}
with parameters $\theta = (\theta_F, \theta_R) \in \mathbb{R}^{2 M}$, the Onsager-Machlup path action Eq.~\eqref{eq:forward-path-action} and the time-asymmetric work Eq.~\eqref{eq:shadow-work} become quadratic in $\theta$

\begin{align}
  S[\trajx(t); \theta] &= \theta^\mu_F \theta^\nu_F \mathsf{a}_{\mu \nu} + \theta^\mu_F \mathsf{b}_{\mu} + \theta\mbox{-}\mathrm{ind. \ terms} \label{eq:quadratic-action}  \\
  W[\trajx(t); \theta] &= - (\theta^\mu_F \theta^\nu_F \mathsf{a}_{\mu \nu} + \theta^\mu_F \mathsf{b}_{\mu} + \mathsf{c}) + \nonumber \\
  &\quad\quad (\theta^\mu_R \theta^\nu_R \mathsf{\tilde{a}}_{\mu \nu} + \theta^\mu_R \mathsf{\tilde{b}}_{\mu} + \mathsf{\tilde{c}}) , \label{eq:quadratic-work}
\end{align}
where

\begin{widetext}
\begin{equation}
    \begin{split}
    \mathsf{a}_{\mu \nu}[\trajx(t)] = (\mathrm{I})& \int_0^{t_f} \frac{\nabla U_\mu \cdot \nabla U_\nu}{4} \mathrm{d}t, \\
    \mathsf{\tilde{a}}_{\mu \nu}[\trajx(t)] = (\mathrm{BI}&) \int_0^{t_f} \frac{\nabla {U}_\mu \cdot \nabla {U}_\nu}{4} \mathrm{d}t,
  \end{split}
    \quad\quad\quad
  \begin{split}
    \mathsf{b}_{\mu}[\trajx(t)]  = \ &(\mathrm{I}) \int_0^{t_f} \frac{\dot{\trajx} \cdot \nabla U_\mu}{2} \mathrm{d}t, \\
    \mathsf{\tilde{b}}_{\mu}[\trajx(t)]  =  - &(\mathrm{BI}) \int_0^{t_f} \frac{\dot{\trajx} \cdot \nabla {U}_\mu}{2} \mathrm{d}t, 
  \end{split}
  \quad\quad
  \begin{split}
      &\mathsf{c}[\trajx(t)] = U_A(\trajx(0)),\\ \\
      \mathrm{and} \quad &\mathsf{\tilde{c}}[\trajx(t)] = U_B(\trajx(t_f)) \label{eq:abc}
  \end{split}
\end{equation}
\end{widetext}
are $\theta$-\textit{independent} functionals of the time-discretized trajectory $\trajx(t)$ \cite{a-tilde-a}. Here, (BI) refers to a Backwards Itô integral, needed to write terms of the reverse ensemble $\tilde{S}[\tilde{\trajx}(t)]$ as a functional of $\trajx(t)$. (Eqs.~\eqref{eq:quadratic-action}--\eqref{eq:abc} apply for the reverse path ensemble $\tilde{S}[\tilde{\trajx}(t)], \tilde{W}[\tilde{\trajx}(t)]$, through the transformation $t \rightarrow t_f - t$, $\{F, R\} \rightarrow \{R, F \}$.) These variables $\mathsf{a}, \mathsf{\tilde{a}} \in \mathbb{R}^{M \times M}, \mathsf{b}, \mathsf{\tilde{b}} \in \mathbb{R}^M$, and $\mathsf{c}, \mathsf{\tilde{c}} \in \mathbb{R}$ are akin to the eligibility trace variables (sometimes called ``Malliavin weights'') used in reinforcement learning policy-gradient algorithms \cite{williams1992simple, peters2008reinforcement, warren2012malliavin, das2019variational, das2022direct}, which are dynamically evolved with each trajectory $\trajx(t)$.

In the following two paragraphs we consider only the forward ensemble for simplicity. If for every trajectory $\trajx^i(t)$ we calculate not only its work $W^i$ but also its functional values $(\mathsf{a}^{i}, \mathsf{b}^{i}, \mathsf{c}^{i}, \mathsf{\tilde{a}}^{i}, \mathsf{\tilde{b}}^{i}, \mathsf{\tilde{c}}^{i}$) and keep track of the $\theta^{i} = (\theta_F^{i}, \theta_R^{i})$ under which it was produced, then a $\theta$-dependent importance-sampling estimator for $\langle W \rangle_F$ may be constructed
\begin{equation}
  \hat{J}_F(\theta) = \frac{\sum_{i=1}^{n_s} \mathsf{r}^i_F (\theta) \, \mathsf{w}^i_F(\theta) }{ \sum_{i=1}^{n_s} \mathsf{r}^i_F(\theta) }, \label{eq:importance-sampling-W-estimator}
\end{equation}
where the sum is over collected forward samples $i$, $\mathsf{r}^i_F(\theta)$ is the likelihood ratio (i.e., the Radon–Nikodym derivative)
\begin{equation}
    \mathsf{r}^i_F(\theta) = \frac{\mathcal{P}[\trajx^i(t) \  \mathrm{from} \ \theta]}{\mathcal{P}[\trajx^i(t) \  \mathrm{from} \  \theta^{i}]} = e^{-\beta (S[ \trajx^i(t); \theta] - S[ \trajx^i(t); \theta^i] )}
\end{equation}
using Eq.~\eqref{eq:quadratic-action}, and $\mathsf{w}^i(\theta) = W[\trajx^i(t); \theta]$ defined in Eq.~\eqref{eq:quadratic-work} is the time-asymmetric work for the trajectory $\trajx^i(t)$ as if it were sampled under $\theta$ instead of $\theta^{i}$. \textit{Protocols may now be optimized by minimizing Eq.~\eqref{eq:importance-sampling-W-estimator} as an objective function.} 

Of course, the quality of the importance-sampling estimate Eq.~\eqref{eq:importance-sampling-W-estimator} degrades the further away the input $\theta$ is from the set of $\theta^{i}$ under which samples were collected. One common heuristic of this degradation is the effective sample size \cite{mcbook}
\begin{equation}
    n_F^{\mathrm{eff}}(\theta) = \frac{ \big(\sum_{i=1}^{n_s} \mathsf{r}^i_F(\theta)\big)^2 }{ \sum_{i=1}^{n_s} \mathsf{r}^i_F(\theta)^2 }, 
\end{equation}
ranging from $1$ (uneven $\mathsf{r}^i_F$ values, high degradation) to $n_s$ (equal $\mathsf{r}^i_F$ values, low degradation).

We now state our algorithm (pseudocode given in the SM \cite{SM}): at each iteration, an equal number of independent forward and reverse trajectories are simulated through Eqs.~\eqref{eq:forward-langevin} and \eqref{eq:reverse-langevin} using the $U_F, U_R$ specified by current parameters $\theta_\mathrm{curr}$, with the time-asymmetric work $W$ and auxiliary variables ($\mathsf{a}, \mathsf{b}, \mathsf{c}, \mathsf{\tilde{a}}, \mathsf{\tilde{b}}, \mathsf{\tilde{c}}$) of each trajectory dynamically calculated; then the protocol is updated through solving the nonlinear constrained optimization problem
\begin{equation}
    \theta_\mathrm{next} = \mathrm{argmin}_\theta \big\{ \hat{J}(\theta) \, | \, \{n_F^{\mathrm{eff}}(\theta),n_R^{\mathrm{eff}}(\theta)\} \geq f n_s \big\}, \label{eq:optimization}
\end{equation}
for which there are efficient numerical solvers (e.g. SLSQP \cite{kraft1988software} pre-implemented in SciPy \cite{2020SciPy-NMeth}). Here $\hat{J}(\theta) = \hat{J}_F(\theta) + \hat{J}_R(\theta)$, $n_F^{\mathrm{eff}}(\theta)$ and $n_R^{\mathrm{eff}}(\theta)$ are constructed with the $n_s$ forward and $n_s$ reverse samples collected over all iterations, and $f \in [0, 1)$ is a hyperparameter specifying the constraint strength: the fraction of total samples we are constraining $n^\mathrm{eff}_{F,R}$ to be greater or equal than. Finally, $\widehat{\Delta F}_\mathrm{BAR}$ is calculated with the bi-directional work measurements collected across all iterations using Eq.~\eqref{eq:BAR-estimator}, which is permitted by the satisfaction of Eq.~\eqref{eq:mixtures}.

\section{Numerical examples}

In this section we report the performance of our algorithm for three test model systems.

We chose our basis set in order to represent protocols of the form
\begin{equation}
    U(\boldx, t) = \lambda_{A}(t) \, U_A(\boldx) + \lambda_{B}(t) \, U_B(\boldx) + \lambda_C(t) \, U_C(\boldx) \label{eq:protocol-form}
 \end{equation}
where $U_C(\boldx)$ is a linear quasi-counterdiabatic potential 
\begin{equation}
    U_C(\boldx) = -\boldc \cdot \boldx \quad\mathrm{with}\quad \boldc = \frac{\langle \boldx \rangle_B - \langle \boldx \rangle_A}{t_f} 
\end{equation}
directly forcing each coordinate $x_n$ with magnitude proportional to the difference of its equilibrium values $\langle \boldx \rangle_{A,B} = \int \boldx \, \rho_{A,B}(\boldx) \, \mathrm{d}\boldx$ \cite{protocol-form}. The basis set is given by
\begin{equation}
    \bigg\{ U_{\ell}(\boldx) \, p_m\bigg(\frac{2t}{t_f} - 1 \bigg)  \, \bigg| \, \ell \in \{A, B, C\}, \, 0 \leq m \leq m_\mathrm{max} \bigg\},  
\end{equation}
where $p_m(\cdot)$ is the $m$-th Legendre polynomial.

For all numerical examples, $m_\mathrm{max} = 4$ was used and the algorithm was initialized with 120 bi-directional samples drawn from a generic naive linear interpolation protocol $\lambda_A(t) = 1 - t/t_f, \lambda_B(t) = t/t_f, \lambda_C(t) = 0$ for both $U_F$ and $U_R$. At each iteration Eq.~\eqref{eq:optimization} was solved for $n_\mathrm{mb} = 20$ independently subsampled minibatches of size $n_s^\mathrm{mb} = 80$ with $f = 0.3$; the protocol was then updated to the minibatch-averaged $\theta_\mathrm{next}$; finally, 20 additional bi-directional samples were drawn with the new protocol. In total, 44 iterations were performed, giving 1000 bi-directional samples.

\begin{figure}[h]
	\centering
	\includegraphics[width=.48\textwidth]{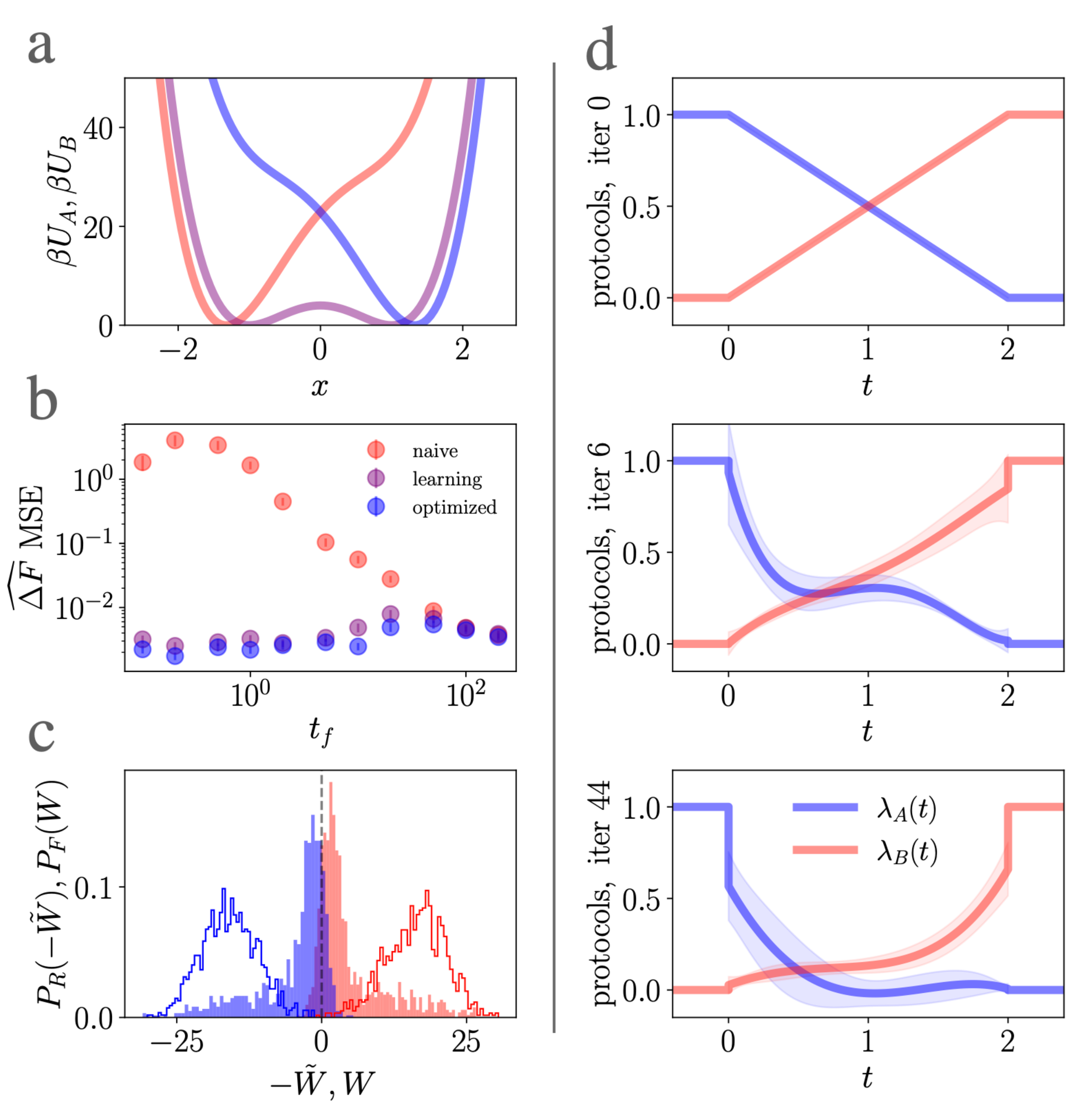}
	\caption{ (a) The potentials $U_A(\boldx)$ (red) and $U_B(\boldx)$ (blue) are obtained by linearly biasing a double well (purple). (b) $\widehat{\Delta F}_\mathrm{BAR}$ mean squared error from 1000 bi-directional measurements drawn solely from the naive protocol (red), cumulatively from protocols that are adaptively optimized (``learning'') with our algorithm (purple), and solely from the last-iteration (``optimized'') protocols (blue) for various protocol times $t_f$. (c) Single-trial $\mathcal{P}_F(W)$ (red) and $\mathcal{P}_R(-\tilde{W})$ (blue) work distributions for 1000 measurements from the naive protocol, (unfilled) and adaptively optimized protocols (filled) for $t_f = 2$, the ground truth shown as a grey dashed line. Measurements made with protocol optimization have significantly more overlap, leading to lower estimator error. (d) Forward protocols $\lambda_A(t)$ (blue) and $\lambda_B(t)$ (red) at various iterations of protocol optimization for $t_f = 2$. Shaded region represents variability across $100$ independent trials. In the optimized last-iteration protocol, $\lambda_A(t) + \lambda_B(t)$ (giving the energy scale) is greatly reduced at intermediate times, while $\lambda_B(t) - \lambda_A(t)$ (giving the linear bias) is time-asymmetrically shifted. The reverse protocols (not shown here) are similar \cite{SM}.}
	\label{dw_mw_fig}
\end{figure}

\subsection{Linearly-Biased double well}

The first system we consider is a one-dimensional quartic double-well with a linear bias (Fig.~\ref{dw_mw_fig}(a)). The potentials are $U_A(x) = E_0[ (x^2 - 1)^2/4 - x]$, $U_B(x) = E_0[ (x^2 - 1)^2/4 + x]$ (cf. \cite{zhong2022limited} for optimal protocols minimizing $\langle W_\mathrm{trad}\rangle_F$). We set $U_C(x) = 0$ because $U_B(x) - U_A(x)$ is already linear in $x$. We use $\beta = 1, E_0 = 16$, and a timestep $\Delta t = 1 \times 10^{-3} \tau$ where $\tau = 1$ is the natural timescale (here the length scale, inverse temperature, and friction coefficient are all unity $\ell = \beta = \gamma = 1$).

Fig.~\ref{dw_mw_fig}(b) displays the $\widehat{\Delta F}_\mathrm{BAR}$ estimator mean squared error for 1000 bi-directional work measurements collected solely from the naive protocol (red), the 1000 measurements collected cumulatively over on-the-fly protocol optimization (purple), and 1000 measurements collected solely from the last-iteration (blue). Each dot represents the empirical average over 100 independent trials. Note that the mean squared error is up to 1600 times lower under protocol optimization compared to under the naive protocol (obtained at $t_f = 0.2$). For $t_f \gtrsim 10$ the algorithm does not converge within the 1000 measurements \cite{SM}, leading to less improvement. Fig.~\ref{dw_mw_fig}(c) shows that bi-directional work measurements collected under the protocol optimization algorithm have significantly more overlap than measurements collected from the naive protocol, leading to reduced estimator error. Fig.~\ref{dw_mw_fig}(d) gives snapshots on how the optimal protocol is adaptively learned.

\begin{figure}[h]
	\centering
	\includegraphics[width=.48\textwidth]{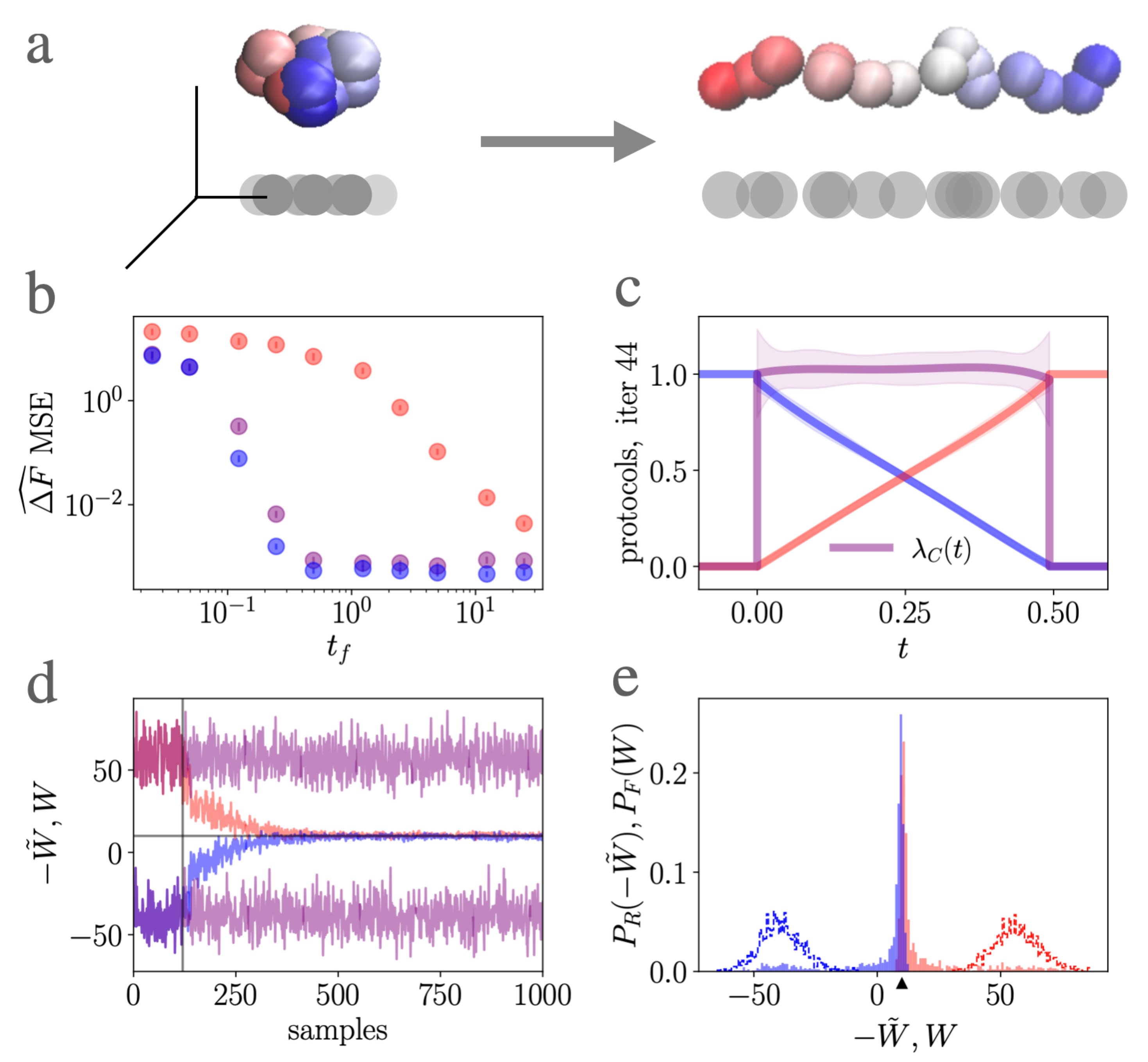}
	\caption{ (a) A Rouse polymer is stretched from a collapsed state to an extended one. (b) $\widehat{\Delta F}_\mathrm{BAR}$ mean squared error verses protocol time, points colored as in Fig.~\ref{dw_mw_fig}. (c) For moderate protocol times ($t_f = 0.5 \, \tau_\mathrm{R}$ displayed here) the optimized protocol $(\lambda_A(t), \lambda_B(t), \lambda_C(t))$ learned in 44 iterations is the counterdiabatic protocol Eq.~\eqref{eq:rouse-cd}. (d) Single-trial bi-directional work samples from the naive protocol ($W$ and $-\tilde{W}$ purple) and adaptively-optimized protocols ($W$ red, $-\tilde{W}$ blue) for $t_f = 0.5 \, \tau_\mathrm{R}$. Vertical line demarcates start of protocol optimization. The ground truth $\Delta F$ is shown as a horizontal line. (e) Work distributions corresponding to the samples in (d). The ground truth is indicated by the triangular arrow. Cumulative measurements made under protocol optimization (filled) have dramatically greater overlap than measurements made under the naive protocols (unfilled), leading to lower estimator error.}
	\label{Rouse_mw_fig}
\end{figure}

\subsection{Rouse polymer}

Next we consider a $(N+1)$-bead Rouse polymer 
(Fig.~\ref{Rouse_mw_fig}(a)) with stiffness $k$ and intrinsic energy given by $U_\mathrm{Rouse}(x_0, x_1, ..., x_{N}) = \sum_{n=0}^{N-1} (k/2) (x_{n+1} - x_n)^2$ from harmonic bonds between adjacent beads \cite{doi1988theory, only-1d}. We estimate $\Delta F$ between a collapsed state (fixing $x_0 = x_N = 0$) and an extended state (fixing $x_0 = 0$, $x_N = \lambda_f$), so our configuration space is $\boldx \in \mathbb{R}^{N - 1}$ with potential energies $U_A(x_1, ..., x_{N-1}) = U_\mathrm{Rouse}(0, x_1, ..., x_{N-1}, 0)$ and $U_B(x_1, ..., x_{N-1}) = U_\mathrm{Rouse}(0, x_1, ..., x_{N-1}, \lambda_f)$. Equilibrium averages $\langle x_n \rangle_A = 0, \langle x_n \rangle_B = n \lambda_f  / N$ give $U_C(\boldx) = - (\lambda_f / N t_f) \sum^N_{n=1}  n x_n$, and the ground truth is $\Delta F = F_B - F_A = k \lambda_f^2 / (2N)$. It may be verified that for this problem the time-varying potential energies

\begin{gather}
  U_0(\boldx, t) = \bigg( 1 - \frac{t}{t_f} \bigg) U_A(\boldx) + \bigg(\frac{t}{t_f} \bigg)U_B(\boldx) \nonumber \\ U_1(\boldx, t) = U_C(\boldx) \label{eq:rouse-cd}
\end{gather}
solve Eq.~\eqref{eq:integrand-reduction} and are thus counterdiabatic. 

We use $\beta  = k = 1$, $N = 20$, and timestep $\Delta t = 2.5 \, \times 10^{-5} \tau_\mathrm{R}$ where $\tau_\mathrm{R} = \beta N^2/\pi^2$ is the Rouse relaxation time \cite{doi1988theory}. Initial conditions for $\rho_A(\boldx)$ and $\rho_B(\boldx)$ were drawn from a normal-modes basis as described in SM Section II \cite{SM}. Fig.~\ref{Rouse_mw_fig}(b) shows an improvement of up to $8300$ (for $t_f = 0.5 \, \tau_\mathrm{R}$) in $\widehat{\Delta F}_\mathrm{BAR}$ mean squared error between naive and optimized protocols. The counterdiabatic solution Eq.~\eqref{eq:rouse-cd} corresponds to $\lambda_A(t) = (1 - t / t_f), \lambda_B(t) = t / t_f$, and $\lambda_C(t) = 1$, which what the algorithm learns for $t_f = \, \tau_\mathrm{R}$ as depicted in Fig.~\ref{Rouse_mw_fig}(c). (This was generally the case for $t_f \geq 0.5 \, \tau_\mathrm{R}$. For $t_f < 0.5 \, \tau_\mathrm{R}$ the algorithm learns a sub-optimal local solution that still provides some improvement \cite{SM}.) Fig.~\ref{Rouse_mw_fig}(d) shows single-trial traces of bi-directional work measurements for the naive protocol (purple) and adaptively-optimized protocols (red for $W$, blue for $-\tilde{W}$), for $t_f = 0.5 \tau_\mathrm{R}$. The protocol converges in $\sim 20$ iterations (requiring $\sim 500$ measurements), and then consistently gives work measurements closely centered at the ground truth free energy (gray horizontal line). Histograms of these traces (filled) are shown in Fig.~\ref{Rouse_mw_fig}(e), exhibiting a remarkable increase in the overlap compared with their naive counterparts (unfilled).

\subsection{Worm-like chain with attractive linker}

We now consider a $(N+1)$-bead worm-like chain model (WLC) in 2 dimensions with an added Lennard-Jones interaction between the first and last beads (similar to the 3rd example of \cite{kuznets2023inferring}). Fixing $(x_0, y_0) = (0, 0)$, the configuration space is $\vec{\phi} \in \mathbb{R}^N$, where $\phi_n$ is the angle of the $n$th bond with respect to the $x$-axis, with $(x_n(\vec{\phi}), y_n(\vec{\phi})) = (\sum_{m=1}^n \cos \phi_m, \sum_{m=1}^n \sin \phi_m)$. The angular potential $U_\phi = k \sum^{N-1}_{n=1} [1 - \cos(\phi_{n+1}-\phi_{n})]$ penalizes the bending of adjacent bonds, and $U_\mathrm{LJ} = 4 \epsilon_\mathrm{LJ} [(\sigma_\mathrm{LJ}/ r_N )^{12}-(\sigma_\mathrm{LJ} / r_N )^6]$ specifies the interaction between first and last beads, where $r_N = \sqrt{ x_N^2 + y_N^2}$ is the end-to-end distance. We take $k = 6, \beta = 1, \epsilon_\mathrm{LJ} = 8, \sigma_\mathrm{LJ} = 4$, and $N = 15$. 

\begin{figure}[h]
	\centering
	\includegraphics[width=.49\textwidth]{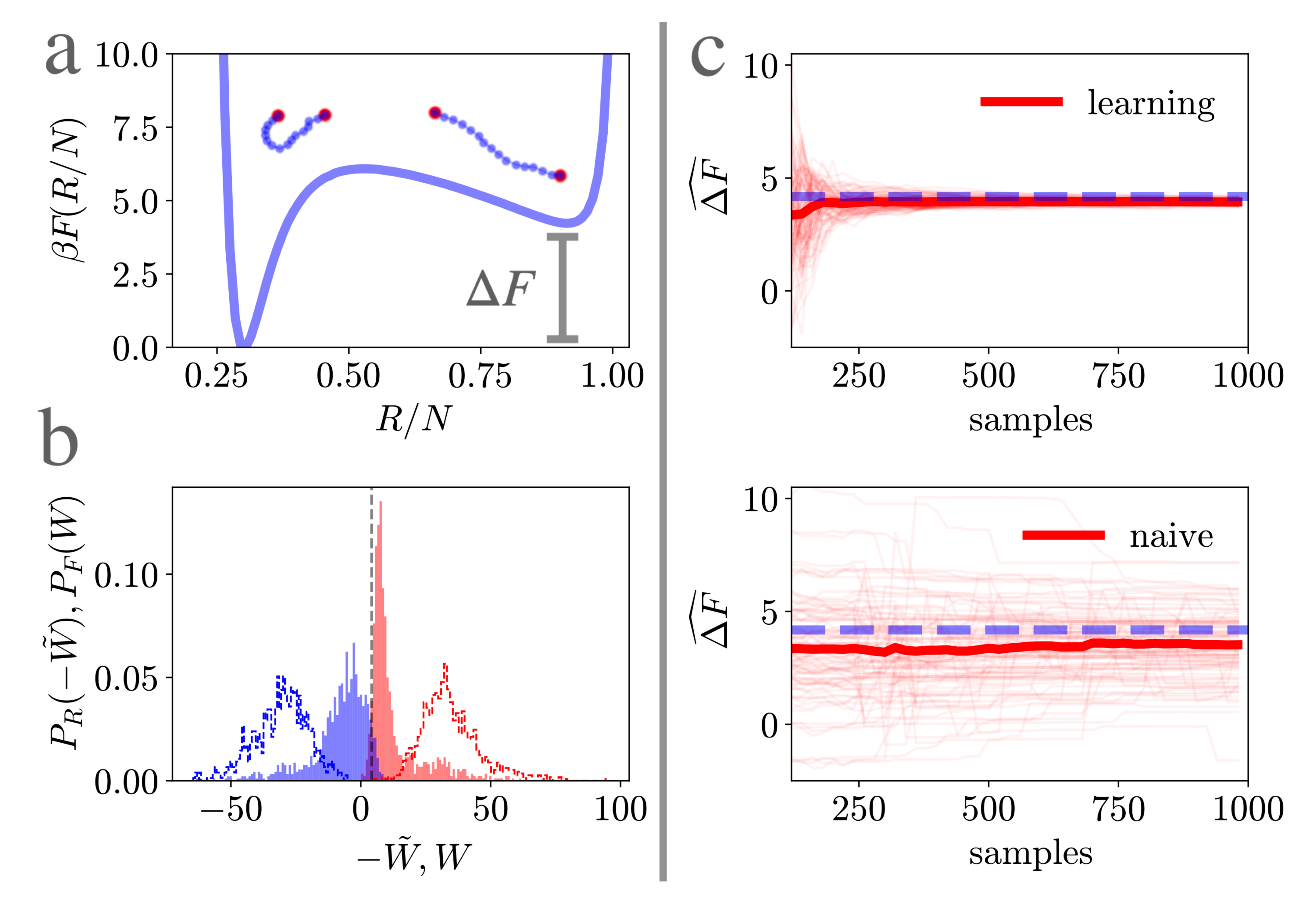}
	\caption{ Worm-like chain with an attractive linker. (a) Ground truth free energy surface relative to its value at $R_A = 2^{1/6} \sigma_\mathrm{LJ}$; the left well corresponds to the ends of the chain bound to one another and the right well corresponds to a nearly straight configuration, with a free energy difference $\Delta F \approx 4.18$. (b) Work distributions before (unfilled) and during (filled) optimization for $t_f = 0.71$, the ground truth shown as a grey dashed line.  (c) The $\widehat{\Delta F}_\mathrm{BAR}$ estimator updated over the $1000$ samples converges to the ground truth value (dashed blue line) much more quickly under protocol optimization than under the naive protocol. At $1000$ samples the protocol optimization free energy estimate was $\widehat{\Delta F} = 3.94 \pm 0.11$, while for the naive protocol was $\widehat{\Delta F} = 3.52 \pm 1.48$ (c.f. ground truth value of $\Delta F = 4.18$). It took $200$ total samples for the mean squared error to drop below $1.00 \, (k_\mathrm{B} T)$ under protocol optimization.}
	\label{WLC_mw_fig}
\end{figure}

Fig.~\ref{WLC_mw_fig}(a) displays the conditioned free energy $F(R/N) := - \beta^{-1} \ln \rho_\mathrm{eq}(r_N = R)$, where $\rho_\mathrm{eq}$ is the equilibrium probability of observing the end-to-end distance under $U = U_\phi + U_\mathrm{LJ}$ \cite{XY-radial-FE} (constructed from $10^7$ equilibrium samples of $U_\phi$, obtained with the Metropolis-adjusted Langevin Algorithm \cite{roberts1996exponential}, that were reweighted by $U_\mathrm{LJ}$). $F(R/N)$ exhibits a deep well for $R_A \approx 2^{1/6} \sigma_\mathrm{LJ}$ (trapped/bent state) and a shallow well at large $R_B \approx 0.9 \, N$ (free/relaxed state), separated by a barrier; their difference in value $\Delta F \approx 4.18$ may be calculated by estimating the $\Delta F$ between $U_A(\vec{\phi}) = U_\phi + U_\mathrm{LJ} + (k_\mathrm{ext}/2)(r_N-\lambda_i)^2$ and $U_B(\vec{\phi}) = U_\phi + U_\mathrm{LJ} + (k_\mathrm{ext}/2)(r_N - \lambda_f)^2$ for $\lambda_i = 2^{1/6}\sigma_\mathrm{LJ}, \lambda_f = 0.9 \, N$, and $k_\mathrm{ext} \gg 1$. 

We calculate the $\Delta F$ between $U_A$ and $U_B$ for $k_\mathrm{ext} = 200$. We use timestep $\Delta t = 1.41 \times 10^{-4} \, \tau_\mathrm{LJ}$ where $\tau_\mathrm{LJ} = \sqrt{\epsilon_\mathrm{LJ}/\sigma^2}$ is the Lennard-Jones timescale. We use $U_C(\vec{\phi}) = -\sum_n c_n r_n$, radially pulling on each individual bead, constructed with $c_n = (\langle r_n \rangle_B - \langle r_n \rangle_A)/t_f$ from equilibrium samples of $\rho_A$ and $\rho_B$. Fig.~\ref{WLC_mw_fig}(b) displays single-trial work histograms for $t_f = 0.71 \, \tau_\mathrm{LJ}$, showing work measurements much closer to the ground truth with protocol optimization, compared to the naive protocol. Fig.~\ref{WLC_mw_fig}(c) shows the updating $\widehat{\Delta F}_\mathrm{BAR}$ estimator over 100 independent trials converges substantially faster to the ground truth. With 1000 bi-directional samples under the naive protocol the mean squared error is $1.62 \ (k_\mathrm{B} T)$; under protocol optimization, only 160 samples (i.e., just after two iterations of protocol optimization) are required to have a smaller mean squared error. Over various $t_f$, the mean squared error is up to 120 times lower under protocol optimization compared to under the naive protocol \cite{SM}.

\section{Discussion}

In this paper, we derived the time-asymmetric microscopic fluctuation theorem for the unconventional work introduced by \cite{vaikuntanathan2008escorted}. We then demonstrated its practical utility for free energy estimation by presenting an adaptive time-asymmetric protocol optimization algorithm, whose effectiveness we illustrated with three toy models of varying complexity. Time-asymmetric protocols have been considered before \cite{li2017shortcuts, li2019stochastic, blaber2020skewed}, but to our knowledge we are the first to use $\widehat{\Delta F}_\mathrm{BAR}$ on bi-directional work measurements from adaptive time-asymmetric protocols. A clear next step is to test our algorithm on more physically realistic systems. This work should be straightforward to implement with JAX-MD \cite{schoenholz2020jax}. In principle our algorithm should work with underdamped dynamics \cite{SM, li2019stochastic}, and it should also be possible to adaptively optimize the protocol time $t_f$ and sampling ratio $n_F/n_R$. Another future direction to pursue is to differentially weight early versus later samples in the estimator to account for differences in the variance of work measurements, as the algorithm more closely approximates a counterdiabatic protocol. 

The fast convergence in our method comes from exploiting of the quadratic structure of the Onsager-Machlup path action to construct $\hat{J}(\theta)$, which allows all samples to be used in each optimization step. Typically the most computationally expensive step in a molecular dynamics simulation is calculating potential energy gradients $\nabla U$ to evolve $\trajx(t)$, which does not need to be repeated to evolve $(\mathsf{a}_{\mu \nu}, \mathsf{b}_\mu, ...)$. A valid critique of our algorithm is that the number of auxiliary variables included with each trajectory scales quadratically with the number of basis functions, becoming prohibitively large when considering, for example, a separate control force on each particle of a many particle system. However, we have shown that a small number of basis functions to represent Eq.~\eqref{eq:protocol-form} already produces a substantial improvement in efficiency for our three examples. That said, it is straightforward to add additional basis functions (cf. Eq. (2) of \cite{naden2014linear}), which may be useful for more complex and realistic systems. It would be interesting to apply recent methods \cite{singh2023variational} to learn the optimal set of additional basis functions, that apply force along specific coordinates: bonds, angles, native contacts and other collective variables to further improve performance for larger scale systems. 

As mentioned at the end of the Introduction, our results were independently derived in~\cite{vargas2023transport} within a machine learning context. It is noteworthy that stochastic thermodynamics has shown to be a useful theoretical framework not only for non-equilibrium statistical physics, but also for machine learning in flow-based diffusion models \cite{sohl2015deep, song2020score, doucet2022score, albergo2023stochastic}. In particular, we recognize significant ties between our work, and that of ``Stochastic Normalizing Flows'' \cite{wu2020stochastic}, wherein authors also consider constructing counterdiabatic protocols under the name ``deterministic invertible functions''. It can be shown that counterdiabatic protocols are perfect stochastic normalizing flows, and they report (after sufficient neural-network training) excellent numerical results for sampling and free energy estimation. The primary difference is that in their work they fix $U_0 = (1 - t/t_f) U_A + (t/t_f) U_B$ and use a neural-network ansatz, whereas here we use an adaptive importance sampling algorithm with a linear spatio-temporal basis ansatz for both $U_0$ and $U_1$. Likewise, we note \cite{bernton2019schr} explores time-asymmetric Markovian processes for sampling, building off of entropy-regularized optimal transport wherein solutions of the continuous-time Schrödinger bridge problem involve asymmetrically-controlled diffusion processes \cite{chen2021stochastic} (see also \cite{de2021diffusion, berner2022optimal}). This is intriguing, as solving for optimal time-\textit{symmetric} protocols has been shown to be equivalent to solving the continuous-time formulation \cite{benamou2000computational} of standard optimal transport \cite{aurell2011optimal, chen2019stochastic, nakazato2021geometrical, zhong2022limited, chennakesavalu2023unified}. In light of all this, we suspect deep theoretical connections between stochastic thermodynamics and machine learning may be further uncovered through the time-asymmetric fluctuation theorem. 

Documented code for this project may be found at \url{https://github.com/adrizhong/dF-protocol-optimization}.

\begin{acknowledgments}
The authors would like to thank Nilesh Tripuraneni, Hunter Akins, Chris Jarzynski, Gavin Crooks, Steve Strong, and the participants of the Les Houches ``Optimal Transport Theory and Applications to Physics'' and Flatiron ``Measure Transport, Diffusion Processes and Sampling'' workshops for useful discussions; Jorge L. Rosa-Raíces for helpful comments on an earlier manuscript version; and Evie Pai for lending personal computing resources. This research used the Savio computational cluster resource provided by the Berkeley Research Computing program at the University of California, Berkeley (supported by the University of California Berkeley Chancellor, Vice Chancellor for Research, and Chief Information Officer). AZ is supported by the Department of Defense (DoD) through the National Defense Science \& Engineering Graduate (NDSEG) Fellowship Program. BKS is supported by the Kavli Energy Nanoscience institute through the Philomathia Foundation Fellowship. MRD thanks Steve Strong and Fenrir LLC for supporting this project. This work was supported in part by the U.S. Army Research Laboratory and the U.S. Army Research Office under contract W911NF-20-1-0151.
\end{acknowledgments}

\appendix

\section{Microscopic fluctuation theorem} \label{appendix:MFT}

\subsection{The Onsager-Machlup Action} \label{appendix:OM-action}

For overdamped Langevin dynamics for $\trajx(t) \in \mathbb{R}^d$, 

\begin{equation}
    \dot{\trajx} = -\nabla U(\trajx(t), t) + \sqrt{2\beta^{-1} } \, \boldeta(t) \quad\mathrm{with}\quad \trajx(0) \sim \rho(\cdot) ,\label{eq:SI-langevin}
\end{equation}
where $\boldeta(t)$ is an instantiation of standard Gaussian white noise with statistics $\langle \eta_i(t) \rangle = 0$ and $\langle \eta_i(t) \eta_j(t') \rangle = \delta_{ij} \delta(t - t')$, and $\rho(\cdot)$ is its initial distribution, the formal expression for the probability of a path's realization is (up to a multiplicative factor) 

\begin{equation}
    \mathcal{P}[\trajx(t)] = \rho(\trajx(0)) e^{-\beta S[\trajx(t)]}. \label{eq:SI-path-probability}
\end{equation}
Here $S[\trajx(t)]$ is the Onsager-Machlup Path Action functional 

\begin{equation}
  S[\trajx(t)] = (\mathrm{I}) \int_0^{t_f} \frac{|\dot{\trajx}(t) + \nabla U(\trajx(t), t) |^2 }{4} \, \mathrm{d} t
\end{equation}
which comes from the path discretization into $N$ timesteps with timestep $\Delta t = t_f / N$: $\trajx(t) \rightarrow [\trajx_0, \trajx_1,  ... \, , \trajx_N]$, with $\trajx_n \approx \trajx(t_n)$, $N = t_f / \Delta t$, and $t_n = n \Delta t$, generated from Euler-Maruyama dynamics

\begin{gather} 
\trajx_0 \sim \rho(\cdot) \\ 
\trajx_{n+1} = \trajx_n - \nabla U(\trajx_n, t_n) \, \Delta t + \sqrt{2 \beta^{-1}} \, \Delta B_n,
\end{gather} 
where $\Delta B_n \sim \mathcal{N}(0, \Delta t \, I_{d})$ is a $d$-dimensional Gaussian random variable (i.e., Brownian increment) \cite{adib2008stochastic}.

The probability of the realization of a particular path is then

\begin{widetext}
  \begin{align}
    \mathcal{P}(\trajx_0, \trajx_1, ... \, , \trajx_N ) &= \mathcal{P}(\trajx_0) \mathcal{P}(\trajx_1 | \trajx_0) \mathcal{P}(\trajx_2 | \trajx_1) ... \mathcal{P}(\trajx_N | \trajx_{N-1}) \nonumber  \\ 
    &= \rho(\trajx_0) \prod_{n = 0}^{N-1} (4 \pi \beta^{-1} \Delta t)^{-d/2} \exp \bigg( -\frac{|\trajx_{n+1} - \trajx_n + \nabla U(\trajx_n, t_n) \Delta t|^2}{4 \beta^{-1} \Delta t} \bigg) \nonumber \\  
    &\propto \rho(\trajx_0) \exp \bigg( - \beta \sum_{n=0}^{N - 1} \frac{ | ( \frac{\trajx_{n+1} - \trajx_n}{\Delta t}) + \nabla U(\trajx_n, t_n) |^2 }{4}\Delta t \bigg),
  \end{align}
\end{widetext}
where the normalization factor $ (4\pi \beta^{-1} \Delta t)^{-Nd/2}$ is hidden in the last line. 

Taking $N \rightarrow \infty$ with $\Delta t = t_f / N \rightarrow 0$, the sum within the exponential becomes

\begin{align}
    &\sum_{n=0}^{N - 1} \frac{ | ( \frac{\trajx_{n+1} - \trajx_n}{\Delta t}) + \nabla U(\trajx_n, t_n) |^2 }{4}\Delta t \nonumber \\ 
    &\quad\longrightarrow\quad (\mathrm{I}) \int_0^{t_f} \frac{|\dot{\trajx}(t) + \nabla U(\trajx(t), t)|^2}{4} \mathrm{d} t = S[\trajx(t)],
\end{align}
which yields the formal expression Eq.~\eqref{eq:SI-path-probability}. 

\subsection{Stochastic Integrals and Itô's formula} \label{appendix:stochastic-integrals}

Here we briefly review the rules of stochastic calculus. For a stochastic path (i.e., a ``Brownian motion'') $\trajx(t)|_{t \in [0, t_f]}$ from Eq.~\eqref{eq:SI-langevin} and some vector-valued function $b(\boldx,t)$, the three following choices for the time-discretization of the integral $\int_0^{t_f} b(\trajx(t), t) \cdot \dot{\trajx}(t) \, \mathrm{d} t$ :

\begin{gather*}
  \sum_{n=0}^{N - 1} b(\trajx_n, t_n) \cdot \Delta \trajx_n, \\
  \sum_{n=0}^{N - 1} b(\trajx_{n + \frac{1}{2}}, t_{n + \frac{1}{2}}) \cdot \Delta \trajx_n, 
\end{gather*}
and 

\begin{align*}
  \sum_{n=0}^{N - 1} b(\trajx_{n+1}, t_{n+1}) \cdot \Delta \trajx_n 
\end{align*}
(here $\Delta \trajx_n = (\trajx_{n+1} - \trajx_n)$, $\trajx_{n + \frac{1}{2}} = (\trajx_n + \trajx_{n+1}) / 2$, and $t_{n + \frac{1}{2}} = (t_n + t_{n+1}) / 2$) do \textit{not} necessarily converge to the same value under the $N \rightarrow \infty$, $\Delta t = t_f / N \rightarrow 0$ limit. This is in contrast to the case where $\trajx(t)|_{t \in [0, t_f]}$ is continuously differentiable, e.g., the solution of a well-behaved \textit{deterministic} differential equation, in which case the above three time-discretizations do converge to the same integral value under the limit \cite{pugh2002real}. 

Therefore, for trajectories $\trajx(t)|_{t \in [0, t_f]}$ obtained through the stochastic differential equation Eq.~\eqref{eq:SI-langevin}, we must define each of these as distinct integrals

\begin{gather*}
   (\mathrm{I}) \int_0^{t_f} b(\trajx(t), t) \cdot \dot{\trajx} \, \mathrm{d} t := \lim_{N \rightarrow \infty} \sum_{n=0}^{N - 1} b(\trajx_n, t_n) \cdot \Delta \trajx_n,  \\
  (\mathrm{S}) \int_0^{t_f} b(\trajx(t), t) \cdot \dot{\trajx} \, \mathrm{d} t := \lim_{N \rightarrow \infty} \sum_{n=0}^{N - 1} b(\trajx_{n + \frac{1}{2}}, t_{n + \frac{1}{2}}) \cdot \Delta \trajx_n, 
\end{gather*}
and 

\begin{gather*}
  (\mathrm{BI}) \int_0^{t_f} b(\trajx(t), t) \cdot \dot{\trajx} \, \mathrm{d} t := \lim_{N \rightarrow \infty} \sum_{n=0}^{N - 1} b(\trajx_{n+1}, t_{n+1}) \cdot \Delta \trajx_n ,
\end{gather*}
which are the Itô, Stratonovich, and Backwards Itô integrals, respectively. They are related to one another by Itô's lemma \cite{oksendal2013stochastic}

\begin{align}
  &(\mathrm{I}) \int_0^{t_f} b(\trajx(t), t) \cdot \dot{\trajx} + \beta \, \nabla \cdot b(\trajx(t), t) \, \mathrm{d} t \nonumber \\
  &\quad= (\mathrm{S}) \int_0^{t_f} b(\trajx(t), t) \cdot \dot{\trajx} \, \mathrm{d} t \nonumber \\
  &\quad\quad= (\mathrm{BI}) \int_0^{t_f} b(\trajx(t), t) \cdot \dot{\trajx} - \beta \, \nabla \cdot b(\trajx(t), t) \, \mathrm{d} t. \label{eq:SI-itos-formula}
\end{align}
The Stratonovich integration convention (i.e., with the time-symmetric midpoint-rule discretization) is particularly convenient because ordinary calculus rules (e.g., the chain rule, product rule, etc.) apply. 

Note that the three separate time-discretizations of integrals of the form $\int_0^{t_f} f(\trajx(t), t) \, \mathrm{d} t$ :

\begin{gather*}
  \sum_{n=0}^{N - 1} f(\trajx_n, t_n) \,\Delta t , \\ 
  \sum_{n=0}^{N - 1} f(\trajx_{n + \frac{1}{2}}, t_{n + \frac{1}{2}}) \, \Delta t ,
\end{gather*}

and 
\begin{align*}
  \quad \sum_{n=0}^{N - 1} f(\trajx_{n+1}, t_{n+1}) \, \Delta t
\end{align*}
\textit{do} converge to the same value under the $N \rightarrow \infty$ with $\Delta t = t_f / N \rightarrow 0$ limit, thus $(\mathrm{I}) \int_0^{t_f} f(\trajx(t), t) \, \mathrm{d} t = (\mathrm{S}) \int_0^{t_f} f(\trajx(t), t) \, \mathrm{d} t = (\mathrm{BI}) \int_0^{t_f} f(\trajx(t), t) \, \mathrm{d} t$.

\subsection{Microscopic fluctuation theorem derivation} \label{appendix:MFT-derivation}

Here, we use the stochastic calculus reviewed above to derive Eq.~\eqref{eq:MFT}, i.e., the equivalence of its first and second lines. We start by manipulating the expression within the exponent

\begin{widetext}
  \begin{gather}
U_A(\trajx(0)) + S[\trajx(t)] + W[\trajx(t)] = U_B(\trajx(t_f)) + (\mathrm{I}) \int_0^{t_f} \bigg\{ \frac{|\dot{\trajx} + \nabla(U_0 + U_1)|^2}{4} - (\dot{\trajx} + \nabla U_1) \cdot \nabla U_0 + \beta^{-1} \nabla^2 (U_1 - U_0) \bigg\} \, \mathrm{d} t  \nonumber \\ 
= U_B(\trajx(t_f)) + \frac{1}{4} (\mathrm{I}) \int_0^{t_f} |\dot{\trajx}|^2 \, \mathrm{d} t + \frac{1}{4} (\mathrm{I}) \int_0^{t_f} |\nabla (U_1 - U_0 ) |^2 \, \mathrm{d} t + \frac{1}{2} (\mathrm{I}) \int_0^{t_f} \dot{\trajx} \cdot \nabla (U_1 - U_0 ) + 2 \beta \nabla^2 (U_1 - U_0) \, \mathrm{d} t \nonumber  \\
= U_B(\trajx(t_f)) + \frac{1}{4} (\mathrm{BI}) \int_0^{t_f} |\dot{\trajx}|^2 \, \mathrm{d} t + \frac{1}{4} (\mathrm{BI}) \int_0^{t_f} |\nabla (U_1 - U_0 ) |^2 \, \mathrm{d} t + \frac{1}{2} (\mathrm{BI}) \int_0^{t_f} \dot{\trajx} \cdot \nabla (U_1 - U_0 )  \, \mathrm{d} t  \nonumber  \\
= U_B(\trajx(t_f)) + (\mathrm{BI}) \int_0^{t_f} \frac{|\dot{\trajx} + \nabla (U_1 - U_0) |^2}{4} \, \mathrm{d} t \nonumber  \\
= U_B(\tilde{\trajx}(0)) + (\mathrm{I}) \int_0^{t_f} \frac{|-\dot{\tilde{\trajx}} + \nabla (\tilde{U}_1 - \tilde{U}_0) |^2}{4} \, \mathrm{d} t \nonumber  \\
= U_B(\tilde{\trajx}(0)) + \tilde{S}[\tilde{\trajx}(t)], \quad\mathrm{where}\quad \tilde{S}[\tilde{\trajx}(t)] = (\mathrm{I}) \int_0^{t_f} \frac{|\dot{\tilde{\trajx}} + \nabla (\tilde{U}_0 - \tilde{U}_1) |^2}{4} \, \mathrm{d} t. 
\end{gather}
\end{widetext}
The first equality comes from using Ito's lemma $U_0(X(t_f), t_f) - U_0(X(0), 0) = (\mathrm{I}) \int_0^{t_f} \partial_t U_0 + \dot{X} \cdot \nabla U_0 \, + \beta^{-1} \nabla^2 U_0 \, \mathrm{d} t$. The second equality follows from standard algebraic manipulation. The third equality comes from converting the Forward Itô integrals $(\mathrm{I})$ to Backward Itô integrals $(\mathrm{B I})$ using Itô's formula Eq.~\eqref{eq:SI-itos-formula}. The fourth equality results from standard algebraic manipulation. The fifth equality comes from the time-reversal transformation $t \rightarrow t_f - t$, with the Backward Itô integral becoming a forward Itô integral under time-reversal.

Finally, we plug in the above to the first line of Eq.~\eqref{eq:MFT} to obtain

\begin{align}
    \tilde{\mathcal{P}}[\tilde{\trajx}(t)] &= e^{-\beta \{ U_A(\trajx(0)) - F_A + S[\trajx(t)] + W[\trajx(t)] - \Delta F\} } \nonumber \\ 
    &= e^{-\beta\{ U_B (\tilde{\trajx}(0)) - F_B + \tilde{S}[\tilde{\trajx}(t)] \} } \nonumber \\ 
    &= \rho_B(\tilde{\trajx}(0)) e^{-\beta \tilde{S}[\tilde{\trajx}(t)]  } ,
\end{align}
thus completing our derivation.

\section{Deriving the fluctuation theorem from the microscopic fluctuation theorem} \label{appendix:MFT-FT}

In this section we derive the Crooks Fluctuation Theorem

\begin{equation}
    \frac{\mathcal{P}_F(+W)}{\mathcal{P}_R(-W)} = e^{+\beta(W - \Delta F)} 
\end{equation}
from the microscopic fluctuation theorem

\begin{equation}
    \mathcal{P}[\trajx(t)] e^{-\beta W[\trajx(t)]} = \tilde{\mathcal{P}}[\tilde{\trajx}(t)] e^{-\beta \Delta F}.\label{eq:SI-MFT-2}
\end{equation}

We begin by recalling that $\mathcal{P}_F(\cdot)$, giving the probability of observing a particular work value in the forward ensemble, is defined as 

\begin{align}
    \mathcal{P}_F(w) &= \langle \delta(W - w) \rangle_F \nonumber \\
    &= \int D \trajx(t) \mathcal{P}[\trajx(t)] \delta(W[\trajx(t)] - w),
\end{align}
where we write $w$ to distinguish the argument of $\mathcal{P}_F(\cdot)$ from the path-functional work $W = W[\trajx(t)]$. Here $D \trajx(t)$ denotes an integral over all paths $\trajx(t)|_{t\in[0, t_f]}$, $\mathcal{P}[\trajx(t)]$ is the probability of its realization, and $\delta(\cdot)$ is the Dirac-delta function. Plugging in Eq.~\eqref{eq:SI-MFT-2} to the above expression, we get 

\begin{align}
    \mathcal{P}_F(w) &= \int D \trajx(t) \tilde{\mathcal{P}}[\tilde{\trajx}(t)] e^{+\beta\{W[\trajx(t)] - \Delta F \} }\delta(W[\trajx(t)] - w) \nonumber \\ 
    &=  e^{+\beta (w - \Delta F) }  \int D \trajx(t) \tilde{\mathcal{P}}[\tilde{\trajx}(t)] \delta(W[\trajx(t)] - w)  \nonumber \\ 
    &= e^{+\beta (w - \Delta F) }  \int D\tilde{\trajx}(t) \tilde{\mathcal{P}}[\tilde{\trajx}(t)] \delta(-\tilde{W}[\tilde{\trajx}(t)] - w)  \nonumber \\ 
    &= e^{+\beta(w - \Delta F)} \mathcal{P}_R(-w),
\end{align}
where in the second line we pull out the exponential using the Dirac delta function, in the third line we consider the coordinate change $\trajx(t) \rightarrow \tilde{\trajx}(t)$ (using also $\tilde{W}[\tilde{\trajx}(t)] = -W[\trajx(t)]$, see Eq.~\eqref{eq:reverse-work} and the text that follows it in the main article), 
and in the fourth line we have recognized that the path integral expression is equivalent to the probability of observing the work value $-w$ 
in the reverse path ensemble.

\bibliography{main}

\begin{thebibliography}{62}
\bibitem [{tra()}]{trajectorynotation}%
  \BibitemOpen
  \href@noop {} {}\bibinfo {note} {In this paper, to avoid confusion we use lower case $x$ to denote configurations, and upper case $X(t)$ to denote (stochastic) trajectories.}\BibitemShut {Stop}%
\bibitem [{gam()}]{gamma1}%
  \BibitemOpen
  \href@noop {} {}\bibinfo {note} {Typically there is a factor of the friction coefficient $\gamma$ multiplying the left side of \eqref{eq:forward-langevin}. For notational simplicity we have set it to one, which given that it is equal for all dimensions may always be done through a time-rescaling.}\BibitemShut {Stop}%
\bibitem [{a-t()}]{a-tilde-a}%
  \BibitemOpen
  \href@noop {} {}\bibinfo {note} {The astute reader might point out that under the rules of stochastic calculus, how they are written out $\mathsf{a} = \mathsf{\tilde{a}}$ for all $\boldx(t)$. However, they are different under a finite time-discretization, and thus must both be kept track.}\BibitemShut {Stop}%
\bibitem [{pro()}]{protocol-form}%
  \BibitemOpen
  \href@noop {} {}\bibinfo {note} {The first two terms are a generalization of the protocol form $U_\lambda(\boldx) = (1 - \lambda) \, U_A(\boldx) + \lambda \, U_B(\boldx)$ commonly considered in stochastic thermodynamics. Typically, linear combinations of just $U_A(\boldx)$ and $U_B(\boldx)$ are not effective in transporting $\rho_A(\boldx) \rightarrow \rho_B(\boldx)$, motivating our inclusion of the linear potential $U_C(\boldx)$.}\BibitemShut {Stop}%
\bibitem [{onl()}]{only-1d}%
  \BibitemOpen
  \href@noop {} {}\bibinfo {note} {Because of the harmonic nature of the interbead potential, the dynamics of the Rouse polymer separates in its 3 spatial dimensions; without loss of generality, we can consider just a single spatial dimension.}\BibitemShut {Stop}%
\bibitem [{XY-()}]{XY-radial-FE}%
  \BibitemOpen
  \href@noop {} {}\bibinfo {note} {From the change of coordinates $R = \sqrt{x_N^2 + y_N^2}, \vartheta = \tan^{-1}(y_N / x_N)$ and the radial symmetry of $U_\phi$ and $U_\mathrm{LJ}$, one has $\rho_\mathrm{eq}(R) \propto (2\pi R)^{-1} \rho_\mathrm{eq}(x_N = R, y_N = 0)$.}\BibitemShut {Stop}%
\end{thebibliography}

\end{document}


\newcommand{\tee}{{\mathrm{t}}}
\newcommand{\tf}{{\mathrm{t_f}}}
\newcommand{\A}{{\mathrm{A}}}
\newcommand{\B}{{\mathrm{B}}}
\newcommand{\C}{{\mathrm{C}}}
\newcommand{\D}{{\mathrm{D}}}
\newcommand{\U}{{\mathrm{U}}}
\newcommand{\W}{{\mathrm{W}}}
\newcommand{\boldU}{{\boldsymbol{U}}}
\newcommand{\boldpi}{{\boldsymbol{\pi}}}
\newcommand{\boldpsi}{{\boldsymbol{\psi}}}
\newcommand{\curlyL}{{\cal{L}}}
\newcommand{\curlyW}{{\cal{W}}}

\newcommand{\boldeta}{{\boldsymbol{\eta}}}
\newcommand{\boldlambda}{{\boldsymbol{\lambda}}}
\newcommand{\boldx}{{\boldsymbol{x}}}
\newcommand{\boldc}{{\boldsymbol{c}}}
\newcommand{\bolddx}{{\Delta \boldsymbol{x}}}

\newcommand{\dX}{{\mathrm{d}X}}
\newcommand{\dP}{{\mathrm{d}P}}
\newcommand{\dXtilde}{{\mathrm{d}\tilde{X}}}
\newcommand{\dt}{{\mathrm{d}t}}
\newcommand{\dtau}{{\mathrm{d}\tau}}

\newcommand{\trajx}{X}
\newcommand{\trajp}{P}

\preprint{APS/123-QED}

\affiliation{Department of Physics, University of California, Berkeley, CA, 94720, USA \looseness=-1}
\affiliation{Biophysics Graduate Group, University of California, Berkeley, CA, 94720, USA \looseness=-1}
\affiliation{Redwood Center For Theoretical Neuroscience and Helen Wills Neuroscience Institute, University of California, Berkeley, Berkeley, CA, 94720}

\title{Supplementary Information for ``Time-Asymmetric Protocol Optimization for Efficient Free Energy Estimation''}
\author{Adrianne Zhong}
\affiliation{Department of Physics, University of California, Berkeley, CA, 94720, USA \looseness=-1}
\affiliation{These authors contributed equally \looseness=-1}
\author{Benjamin Kuznets-Speck}
\affiliation{Biophysics Graduate Group, University of California, Berkeley, CA, 94720, USA \looseness=-1}
\affiliation{These authors contributed equally \looseness=-1}
\author{Michael R. DeWeese}
\affiliation{Department of Physics, University of California, Berkeley, CA, 94720, USA \looseness=-1}
\affiliation{Redwood Center For Theoretical Neuroscience and Helen Wills Neuroscience Institute, University of California, Berkeley, Berkeley, CA, 94720}

\date{\today}
\maketitle

\section{Time-asymmetric fluctuation theorem for underdamped dynamics}

In this section, we generalize the time-asymmetric fluctuation theorem to underdamped dynamics. To begin, we review three types of dynamics: overdamped, underdamped, and deterministic. 

At inverse temperature $\beta$, overdamped Langevin dynamics for a stochastic trajectory $X(t) \in \mathbb{R}^d$ are given by the overdamped Langevin equation

\begin{equation}
  \dot{\trajx} = - \mu \nabla_x U_0(X(t),t) + \sqrt{2 \mu \beta^{-1} } \eta(t) 
\end{equation}
where $\mu$ is the mobility, and $\eta(t)$ is an instantiation of standard $d$-dimensional Gaussian white noise, i.e. with statistics $\langle \eta_i(t) \rangle = 0, \langle \eta_i(t) \eta_j(t') \rangle = \delta_{ij} \delta(t - t')$. Note that for generality the mobility is \textit{not} set to one, in contrast to the main text.

On the other hand, underdamped dynamics for position and momentum variables $X(t) \in \mathbb{R}^d, P(t) \in \mathbb{R}^d$ are given by the underdamped Langevin equation

\begin{gather}
  \dot{\trajx} = \frac{P(t)}{m} \nonumber \\ 
  \dot{\trajp} = -\nabla_x U_0(X(t), t) - \gamma P(t) + \sqrt{2 \gamma \beta^{-1} } \zeta(t), 
\end{gather}
where $m$ is the mass, $\gamma$ is the friction coefficient, and $\zeta(t)$ is also an instantiation of standard $d$-dimensional Gaussian white noise with $\langle \zeta_i(t) \rangle = 0, \langle \zeta_i (t) \zeta_j(t') \rangle = \delta_{ij} \delta(t - t')$ \cite{sivak2013using}. Importantly, this underdamped dynamics has the Hamiltonian

\begin{equation}
    H(x, p, t) = \frac{|p|^2}{2m} + U_0(x, t).
\end{equation}
Note that taking the limit $\gamma \rightarrow 0$ reproduces standard Hamiltonian mechanics. 

Finally, we explicitly write out the deterministic dynamics under a flow field $b_1(x, p, t) \in \mathbb{R}^{2d}$ that applies to both position and momentum variables

\begin{align}
  \dot{\trajx} &= b_1^x(X(t), P(t), t) \nonumber \\ 
  \dot{\trajp} &= b_1^p(X(t), P(t), t). 
\end{align}
This vector flow field $b_1$ is a generalization of the time-asymmetric (gradient) force provided by $-\nabla U_1$ in the main text.

\subsection{Generalized Langevin dynamics}

For our derivation, we define a hybridized Langevin equation that combines all of the above three dynamics

\begin{gather}
  \dot{\trajx} = b_1^x(X(t), P(t), t) +  \frac{P(t)}{m} + \big\{ -\mu \nabla_x U_0(X(t),t) + \sqrt{2 \mu \beta^{-1} } \eta(t) \big\} \nonumber \\ 
  \dot{\trajp} = b_1^p(X(t), P(t), t)  -\nabla_x U_0(X(t), t) - \gamma P(t) + \sqrt{2 m \gamma \beta^{-1} } \zeta(t) \nonumber \\ \nonumber \\
  \mathrm{with} \quad\quad X(0), P(0) \sim \rho_A, \label{eq:SI-gen-langevin}
\end{gather}
where $\rho_A$ is the equilibrium distribution corresponding to the Hamiltonian $H_A(x, p) = H(x, p, 0)$ at time $t= 0$, to be specified below. By considering $m$, $\mu$, and $\gamma$ as independent parameters, overdamped dynamics are reproduced by taking the limit $\gamma \rightarrow 0, m \rightarrow \infty$ with the assumption $b_1^x(x, p, t)$ has no $p$-dependence, while underdamped dynamics are reproduced under the limit $\mu \rightarrow 0$, with further taking $\gamma \rightarrow 0$ yielding deterministic Hamiltonian dynamics. 

\subsection{Time-asymmetric work and path action}

In this general setting, we consider protocols that ``switch'' the Hamiltonian between 

\begin{equation}
  H(x, p, 0) = H_A(x, p) = \frac{|p|^2}{2m} + U_A(x) \quad\quad \rightarrow \quad\quad H(x, p, t_f) = H_B(x, p) =\frac{|p|^2}{2m} + U_B(x),
\end{equation}
by switching the potential energy $U(x, 0) = U_A(x) \rightarrow U(x, t_f) = U_B(x)$. The equilibrium distributions for $H_A$ and $H_B$ are given by

\begin{align}
    \rho_A(x, p) &= e^{-\beta [H_A(x, p) - F_A]} \quad \mathrm{with} \quad F_A = -\beta^{-1} \ln \int e^{-\beta H_A(x, p)} \, \mathrm{d}x \, \mathrm{d}p \nonumber \\ 
    \rho_B(x, p) &= e^{-\beta [H_B(x, p) - F_B]} \quad \mathrm{with} \quad F_B = -\beta^{-1} \ln \int e^{-\beta H_B(x, p) } \, \mathrm{d}x \, \mathrm{d}p. 
\end{align}

For ease of notation, from here on out we will denote a phase-space trajectory with variable $Z(t) := (X(t), P(t))$ (not to be confused with the partition function $\exp(-\beta F)$ often denoted by the same variable). 

The unconventional work for a stochastic trajectory $Z(t) |_{t\in [0, t_f]}$ is given by 

\begin{equation}
    W[Z(t)] = \int_0^{t_f} \frac{\partial H}{\partial t} + b_1 \cdot \nabla H - \beta^{-1} \nabla \cdot b_1  \,  \dt, \label{eq:SI-unconventional-work}
\end{equation}
(we use notation $b_1 \cdot \nabla H = b_1^x \cdot \nabla_x H + b_1^p \cdot \nabla_p H$ and $\nabla \cdot b_1 = \nabla_x \cdot b_1^x + \nabla_p \cdot b_1^p$) \cite{vaikuntanathan2008escorted}, while the probability of observing the trajectory under Eq.~\eqref{eq:SI-gen-langevin} is 

\begin{equation}
    \mathcal{P} [Z(t)] = \rho_A(Z(0)) \, e^{- \beta S[Z(t)]} 
\end{equation}
with the path action given by the Itô integral

\begin{equation}
    S[Z(t)] = (\mathrm{I)} \int_0^{t_f} \frac{|\dot{\trajx} - b_1^x - P / m + \mu \nabla_x U_0|^2}{4 \mu} + \frac{|\dot{\trajp} - b_1^p + \nabla_x U_0 + \gamma P |^2}{4 m \gamma } \, \dt. 
\end{equation}
This may be seen by considering the probability of obtaining the particular realization of the noise terms $\eta(t)$ and $\zeta(t)$ that produce the trajectory $Z(t) = (X(t), P(t))$.

\subsection{Derivation of microscopic fluctuation theorem} 

As with the detailed derivation for the overdamped case in Appendix A, we manipulate the sum 

\begin{widetext}
  \begin{gather}
    H_A(Z(0)) + S[Z(t)] + W[Z(t)] \nonumber \\
   = H_B(Z(t_f)) + (\mathrm{I)} \int_0^{t_f} \frac{|\dot{\trajx} - b_1^x - P / m + \mu \nabla_x U_0|^2}{4 \mu} + \frac{|\dot{\trajp } - b_1^p + \nabla_x U_0 + \gamma P |^2}{4 m \gamma } - (\dot{Z} - b_1) \cdot \nabla H  \ -  \nonumber \\ 
    \beta^{-1} (\nabla \cdot b_1 + \mu \nabla^2_x H + m \gamma \nabla^2_p H )  \, \dt  \nonumber
    \\ 
   = H_B(Z(t_f)) + (\mathrm{I}) \int_0^{t_f} \frac{|\dot{X}|^2}{4 \mu} + \frac{|\dot{P}|^2}{4 m \gamma} \, \dt +  
   (\mathrm{I}) \int_0^{t_f} \frac{|-b_1^x - P/m + \mu \nabla_x U_0|^2 }{4 \mu} + \frac{|-b_1^p + \nabla_x U_0 + \gamma P|^2 }{4 m \gamma} \, \dt \ + \nonumber \\ 
   \frac{1}{2} (\mathrm{I}) \int_0^{t_f} \dot{X} \cdot \bigg(  \frac{ -b_1^x - P/m + \mu \nabla_x U_0}{\mu}\bigg) + \dot{P} \cdot \bigg( \frac{-b_1^p + \nabla_x U_0 + \gamma P}{m \gamma} \bigg) \, \dt \ + \nonumber \\ 
   (\mathrm{I}) \int_0^{t_f}  ( b_1^x - \dot{X}) \cdot \nabla_x U_0 + (b_1^p - \dot{P}) \cdot (P / m) - \beta^{-1} ( \nabla_x \cdot (b_1^x +  \mu \nabla_x U_0) + \nabla_p \cdot (b_1^p + m \gamma (P/m)) )  \,  \dt. 
  \end{gather}
\end{widetext}
The first equality comes from plugging in the work and path action definitions and applying the total time derivative with Itô's lemma $H_B(Z(t_f)) - H_A(Z(0)) = (\mathrm{I}) \int_0^{t_f}  \partial_t H + Z \cdot \nabla H + \beta^{-1} (\mu \nabla_x^2 H + m \gamma \nabla_p^2 H) \, \dt$ (here the divergence terms come from the Gaussian white noise on $X(t)$ and $P(t)$, see Appendix A), while the second equality is obtained by expanding out each of the squared terms.

Continuing with our derivation, we insert $\nabla_x \cdot p = \nabla_p \cdot \nabla_x U_0= 0$ into the integral, as well as apply the following two algebraic manipulations

\begin{gather}
    (\mathrm{I}) \int_0^{t_f} \frac{|-b_1^x - P/m + \mu \nabla_x U_0|^2 }{4 \mu} + \frac{|-b_1^p + \nabla_x U_0 + \gamma P|^2 }{4 m \gamma} \, \dt +   (\mathrm{I}) \int_0^{t_f}  b_1^x \cdot \nabla_x U_0 + b_1^p \cdot (P / m) \,  \dt \nonumber \\
    = (\mathrm{I}) \int_0^{t_f} \frac{|-( b_1^x + P/m) + \mu \nabla_x U_0|^2 }{4 \mu} + (b_1^x + (P/m)) \cdot \nabla_x U_0 - (\nabla_x U_0  - b_1^p) \cdot (P/m) + \frac{|(-b_1^p + \nabla_x U_0) + \gamma P|^2 }{4 m \gamma} \, \dt  \nonumber \\ 
    = (\mathrm{I}) \int_0^{t_f} \frac{|b_1^x + P/m + \mu \nabla_x U_0|^2 }{4 \mu} + \frac{|-b_1^p + \nabla_x U_0 - \gamma P|^2 }{4 m \gamma} \, \dt, 
\end{gather}
and 

\begin{gather}
    \frac{1}{2} (\mathrm{I}) \int_0^{t_f} \dot{X} \cdot \bigg(  \frac{ -b_1^x - P/m + \mu \nabla_x U_0}{\mu}\bigg) + \dot{P} \cdot \bigg( \frac{-b_1^p + \nabla_x U_0 + \gamma P}{m \gamma} \bigg) \, \dt - (\mathrm{I}) \int_0^{t_f} \dot{X} \cdot \nabla_x U_0 + \dot{P} \cdot (P/m) \, \dt \nonumber \\ 
    = \frac{1}{2} (\mathrm{I}) \int_0^{t_f} (- \dot{X}) \cdot \bigg(  \frac{ b_1^x + P/m + \mu \nabla_x U_0}{\mu}\bigg) + \dot{P} \cdot \bigg( \frac{-b_1^p + \nabla_x U_0 - \gamma P}{m \gamma} \bigg) \, \dt, 
\end{gather}
ultimately yielding

\begin{widetext}
  \begin{gather}
  H_A(Z(0)) + S[Z(t)] + W[Z(t)] \nonumber 
  \\
    = H_B(Z(t_f)) + (\mathrm{I}) \int_0^{t_f} \frac{|\dot{X}|^2}{4 \mu} + \frac{|\dot{P}|^2}{4 m \gamma} \, \dt + 
    (\mathrm{I}) \int_0^{t_f} \frac{|b_1^x + P/m + \mu \nabla_x U_0|^2 }{4 \mu} + \frac{|-b_1^p + \nabla_x U_0 - \gamma P|^2 }{4 m \gamma} \, \dt \ + \nonumber \\ 
    \frac{1}{2} (\mathrm{I}) \int_0^{t_f} (- \dot{X}) \cdot \bigg(  \frac{ b_1^x + P/m + \mu \nabla_x U_0}{\mu}\bigg) + \dot{P} \cdot \bigg( \frac{-b_1^p + \nabla_x U_0 - \gamma P}{m \gamma} \bigg) \ - \nonumber \\
   2 \beta^{-1}  \bigg\{  \mu \nabla_x \cdot \bigg( \frac{b_1^x + \mu \nabla_x U_0 + P/m }{\mu} \bigg) - m \gamma \nabla_p \cdot \bigg(\frac{ -b_1^p - \gamma P + \nabla_x U_0 }{m \gamma} \bigg) \bigg\}  \,  \dt 
   \\ 
   = H_B(Z(t_f)) + (\mathrm{BI}) \int_0^{t_f} \frac{|\dot{X}|^2}{4 \mu} + \frac{|\dot{P}|^2}{4 m \gamma} \, \dt + 
    (\mathrm{BI}) \int_0^{t_f} \frac{|b_1^x + P/m + \mu \nabla_x U_0|^2 }{4 \mu} + \frac{|-b_1^p + \nabla_x U_0 - \gamma P|^2 }{4 m \gamma} \, \dt \ + \nonumber \\ 
    \frac{1}{2} (\mathrm{BI}) \int_0^{t_f} (-\dot{X}) \cdot \bigg(  \frac{ b_1^x + P/m + \mu \nabla_x U_0}{\mu}\bigg) + \dot{P} \cdot \bigg( \frac{-b_1^p + \nabla_x U_0 - \gamma P}{m \gamma} \bigg) \, \dt  \nonumber 
    \\
    = H_B(Z(t_f)) + (\mathrm{BI}) \int_0^{t_f} \frac{|-\dot{\trajx} + b_1^x + P / m + \mu \nabla_x U_0|^2}{4 \mu} + \frac{|\dot{\trajp } - b_1^p + \nabla_x U_0 - \gamma P |^2}{4 m \gamma } \, \dt
    \\ 
    = H_B(\tilde{Z}(0)) + (\mathrm{I}) \int_0^{t_f} \frac{|\dot{\tilde{\trajx}} + \tilde{b}_1^x - \tilde{P} / m + \mu \nabla_x \tilde{U}_0|^2}{4 \mu} + \frac{|\dot{\tilde{\trajp}} + \tilde{b}_1^p + \nabla_x \tilde{U}_0 + \gamma \tilde{P} |^2}{4 m \gamma } \, \dt =: H_B(\tilde{Z}(0)) + \tilde{S}[\tilde{Z}(t)]. 
  \end{gather}
\end{widetext}
Here, in the second equality we use Itô's lemma to transform from forward to backwards Itô integrals; in the third equality we express the terms back into the squared expression; and in the fourth equality we perform the time-reversal change of variables $\tilde{X}(t) = X(t_f - t), \tilde{P}(t) = -P(t_f - t)$ with  $\tilde{b}^x_1(x, p, t) = b^x_1(x, -p, t_f - t), \tilde{U}_0(x, t) = U_0(x, t_f - t)$, and $\tilde{b}^p_1(x, p, t) = - b^p_1(x, -p, t_f - t)$, which transforms the backwards Itô integral back into a forward Itô integral. 

Here, $\tilde{S}[\tilde{Z}(t)]$ is the path action for the generalized Langevin dynamics 

\begin{gather}
  \dot{\tilde{\trajx}} = -\tilde{b}_1^x(\tilde{X}(t), \tilde{P}(t), t) +  \frac{\tilde{P}(t)}{m} + \big\{ -\mu \nabla_x \tilde{U}_0(\tilde{X}(t),t) + \sqrt{2 \mu \beta^{-1} } \eta(t) \big\} \nonumber \\ 
  \dot{\tilde{\trajp}} = -\tilde{b}_1^p(\tilde{X}(t), \tilde{P}(t), t)   -\nabla_x \tilde{U}_0(\tilde{X}(t), t) - \gamma \tilde P(t) + \sqrt{2 m \gamma \beta^{-1} } \zeta(t), 
\end{gather}
which from inspection differs from Eq.~\eqref{eq:SI-gen-langevin} by having minus signs in front of the $b_1^x$ and $b_1^p$ terms. After specifying the initial conditions $\tilde X(0),  \tilde P(0) \sim \rho_B$, we have a path ensemble for which the probability of observing a particular trajectory $\tilde{Z}(t)|_{t=0}^{t_f}$ satisfies 

\begin{align}
    \tilde{\mathcal{P}}[\tilde{Z}(t)] &= \rho_B(\tilde{Z}(0)) e^{-\beta \tilde{S}[\tilde{Z}(t)]} \nonumber \\
    &= e^{-\beta\{ H_B(\tilde{Z}(0)) - F_B + \tilde{S}[\tilde{Z}(t)] \} } \nonumber \\
    &= e^{-\beta\{ H_A(Z(0)) + S[Z(t)] + W[Z(t)] - F_B \} } \nonumber \\
    &= \rho_A(Z(0)) e^{-\beta \{ S[Z(t)] + F_A - F_B + W[Z(t)] \} } \nonumber \\ 
    &= P[Z(t)] e^{ -\beta\{ W[Z(t)] - \Delta F \} }, 
\end{align}
namely the time-asymmetric microscopic fluctuation theorem for the generalized dynamics Eq.~\eqref{eq:SI-gen-langevin}.

\section{Initial samples for Rouse polymer from normal-modes decomposition} \label{appendix:normal-modes}

For the Rouse polymer, random samples may be drawn by exploiting a normal-modes decomposition. We can write 

\begin{widetext}
    \begin{equation}
      U_{A,B}(x_1, x_2, ..., x_{N-1}) = \bar{U} \bigg(x_1 - \frac{\lambda_{i,f}}{N} , x_2 - \frac{2 \lambda_{i,f}}{N}, ..., x_{N-1} - \frac{(N - 1)\lambda_{i,f}}{N} \bigg) + \frac{k \lambda_{i,f}^2}{2N},
    \end{equation}
\end{widetext}
where $\bar{U}(\vec{y}) = \vec{y}^T K \vec{y} / 2$ with

\begin{equation}
K = 
 \begin{pmatrix} 
    2k & -k & & & &  \\
    -k & 2k &  -k & & \huge{0} &\\
       & -k & & \ddots & & \\
       &    &  \ddots & & -k &  \\ 
       &  \huge{0}  &   & -k & 2k & -k \\
       & &  & & -k & 2k
    \end{pmatrix}.
\end{equation}

Then, we can do an eigenmode decomposition of $K$, writing

$$
\bar{U}(\vec{y}) = \hat{U}(\vec{z}) = \sum_{n=1}^{N-1} \frac{\kappa_n z_n^2}{2},
$$
where 

\begin{align}
\kappa_n &= 2\bigg[ 1 - \cos\bigg( \frac{\pi n }{N }\bigg) \bigg], \\
z_n &= \sqrt{\frac{2}{N}} \, \sum_{m=1}^{N-1} \sin\bigg( \frac{2 \pi n m}{N} \bigg) y_m.
\end{align}

Finally, for an individual initial condition, we draw the normal random variable $z_n \sim \mathcal{N}(\mu = 0, \sigma^2 = (\beta \kappa_n)^{-1})$ for each $n$, as $\hat{\rho}(\vec{z}) \propto \prod \exp (-\beta \kappa_n z_n^2 / 2)$; then we convert from $\vec{z}$ to $\vec{y}$ coordinates via 

\begin{equation}
    y_n = \sqrt{\frac{2}{N}} \, \sum_{m=1}^{N-1}  \sin\bigg( \frac{2 \pi n m}{N} \bigg) z_m,
\end{equation}
before finally adding $x_n = y_n + n \lambda_{i,f} /N$ to get our initial condition. 

Incidentally, the $k \lambda_{i,f}^2 / 2N$ in the expression comparing $U_{A,B}(\vec{x})$ to $U(\vec{y})$ is, up to an additive constant, the free energy $F_{A,B}$.

\begin{figure}
\begin{algorithm}[H]
\caption{Time-Asymmetric Protocol Optimization via Adaptive Importance Sampling}\label{alg:algorithm}
\begin{algorithmic}[1]
  \State \textbf{inputs} $\beta$, $U_A(\boldx)$, $U_B(\boldx)$; stepsize $\mathrm{d}t$, number of timesteps $N$, basis functions $\{U_\mu(\boldx, n)|_{n = 1, .., N}\}$, initial guess $\theta_\mathrm{init}$
  \State \textbf{parameters} Samples per iteration $N_\mathrm{s}$, minibatches per iteration $N_\mathrm{mb}$, minibatch size $n^\mathrm{mb}_s$, constraint strength $f$
  \State \textbf{given} Methods $\Call{DrawSampleA}$, $\Call{DrawSampleB}$ that return equilibrium samples from $\rho_A$, $\rho_B$
  \State \textbf{output} Iteratively updated $\widehat{\Delta F}$ estimate
  \State 
  \Function{RunTrajF}{parameters $\theta = (\theta_F, \theta_R)$} \hfill \Comment{Euler-Maruyama method}
    \State Obtain $\boldx_0 \gets \Call{DrawSampleA}$
    \State Initialize $\boldx, \mathsf{a}_{\mu\nu}, \mathsf{b}_{\mu}, \mathsf{\tilde{a}}_{\mu\nu}, \mathsf{\tilde{b}}_{\mu} \gets \boldx_0, 0, 0, 0, 0$
    \For {$n = 1, ..., N$}
      \State Evaluate $\nabla \mathsf{U}_\mu \gets \nabla U_\mu(\boldx, n) $ for each $\mu$
      \State Calculate $\mathrm{d}\boldx \gets -\theta^\mu_F \nabla \mathsf{U}_\mu \mathrm{d}t + \sqrt{2 \beta^{-1}} \, \mathrm{d} \mathsf{B}$, where $\mathrm{d} \mathsf{B} \sim \mathcal{N}(0, \mathrm{d}t \times I_{d})$ is a $d$-dimensional normal random variable
      \State Evaluate $\nabla \mathsf{\tilde{U}}_\mu \gets \nabla U_\mu(\boldx + \mathrm{d}\boldx, n) $ for each $\mu$
      \State Evolve $\boldx \gets \boldx + \mathrm{d} \boldx$
      \State Evolve $\mathsf{a}_{\mu\nu}, \mathsf{b}_{\mu} \gets \mathsf{a}_{\mu\nu} +  \nabla \mathsf{U}_\mu \cdot \nabla \mathsf{U}_\nu \, \mathrm{d} t / 4,  \mathsf{b}_{\mu} + \nabla \mathsf{U}_\mu \cdot \mathrm{d} \boldx / 2$
      \State Evolve $\mathsf{\tilde{a}}_{\mu\nu}, \mathsf{\tilde{b}}_{\mu} \gets 
      \mathsf{\tilde{a}}_{\mu\nu} +  \nabla \mathsf{\tilde{U}}_\mu \cdot \nabla \mathsf{\tilde{U}}_\nu \, \mathrm{d} t / 4,   \mathsf{\tilde{b}}_{\mu} - \nabla \mathsf{\tilde{U}}_\mu \cdot \mathrm{d} \boldx / 2$ \hfill \Comment{This holds because $\mathrm{d}\tilde{\boldx} = - \mathrm{d} \boldx$}
    \EndFor
    \State Evaluate $\mathsf{c}, \mathsf{\tilde{c}} \gets U_A(\boldx_0), U_B(\boldx)$
    \State Calculate $W \gets - (\theta^\mu_F \theta^\nu_F \mathsf{a}_{\mu\nu} + \theta^\mu_F \mathsf{b}_{\mu} +  \mathsf{c} ) + (\theta^\mu_R \theta^\nu_R \mathsf{\tilde{a}}_{\mu\nu} + \theta^\mu_R \mathsf{\tilde{b}}_{\mu}  +  \mathsf{\tilde{c}})$
    \State \Return $W, \mathsf{a}_{\mu \nu}, \mathsf{b}_\mu, \mathsf{c}, \mathsf{\tilde{a}}_{\mu \nu}, \mathsf{\tilde{b}}_{\mu}, \mathsf{\tilde{c}}, \theta$
  \EndFunction
  \State 
  \Function{RunTrajR}{parameters $\theta = (\theta_F, \theta_R)$} 
    \State Obtain $\tilde{\boldx}_0 \gets \Call{DrawSampleB}$
    \State Initialize $\tilde{\boldx}, \mathsf{\tilde{a}}_{\mu\nu}, \mathsf{\tilde{b}}_{\mu}, \mathsf{a}_{\mu\nu}, \mathsf{b}_{\mu} \gets \tilde{\boldx}_0, 0, 0, 0, 0$
    \For {$n = 1, ..., N$}
      \State Evaluate $\nabla \mathsf{\tilde{U}}_\mu \gets \nabla U_\mu(\tilde{\boldx}, N + 1 - n) $ for each $\mu$ \hfill \Comment{because $\nabla \tilde{U}_\mu(\cdot, n) = \nabla U_\mu(\cdot, N + 1 - n)$} 
      \State Calculate $\mathrm{d}\tilde{\boldx} \gets -\theta^\mu_R \nabla \mathsf{\tilde{U}}_\mu \mathrm{d}t + \sqrt{2 \beta^{-1} } \, \mathrm{d} \mathsf{B}$, where $\mathrm{d} \mathsf{B} \sim \mathcal{N}(0, \mathrm{d}t \times I_{d})$ is a $d$-dimensional normal random variable
      \State Evaluate $\nabla \mathsf{{U}}_\mu \gets \nabla U_\mu(\tilde{\boldx} + \mathrm{d}\tilde{\boldx}, N + 1 - n)$ for each $\mu$
      \State Evolve $\tilde{\boldx} \gets \tilde{\boldx} + \mathrm{d} \tilde{\boldx}$
      \State Evolve $\mathsf{\tilde{a}}_{\mu\nu}, \mathsf{\tilde{b}}_{\mu} \gets \mathsf{\tilde{a}}_{\mu\nu} +  \nabla \mathsf{\tilde{U}}_\mu \cdot \nabla \mathsf{\tilde{U}}_\nu \, \mathrm{d} t / 4,   \mathsf{\tilde{b}}_{\mu} + \nabla \mathsf{\tilde{U}}_\mu \cdot \mathrm{d} \tilde{\boldx} / 2$
      \State Evolve $\mathsf{a}_{\mu\nu}, \mathsf{b}_{\mu} \gets \mathsf{a}_{\mu\nu} +  \nabla \mathsf{U}_\mu \cdot \nabla \mathsf{U}_\nu \, \mathrm{d} t / 4,  \mathsf{b}_{\mu} - \nabla \mathsf{U}_\mu \cdot \mathrm{d} \tilde{\boldx} / 2$
    \EndFor
    \State Evaluate $\mathsf{\tilde{c}}, \mathsf{c} \gets U_B(\tilde{\boldx}_0), U_A(\tilde{\boldx})$
    \State Calculate ${\tilde{W}} \gets - (\theta^\mu_R \theta^\nu_R \mathsf{\tilde{a}}_{\mu\nu} + \theta^\mu_R \mathsf{\tilde{b}}_{\mu}  +  \mathsf{\tilde{c}}) + (\theta^\mu_F \theta^\nu_F \mathsf{a}_{\mu\nu} + \theta^\mu_F \mathsf{b}_{\mu} +  \mathsf{c} ) $
    \State \Return  ${\tilde{W}}, \mathsf{\tilde{a}}_{\mu \nu}, \mathsf{\tilde{b}}_\mu, \mathsf{\tilde{c}}, \mathsf{a}_{\mu \nu}, \mathsf{b}_\mu, \mathsf{c}, \theta$
  \EndFunction

  \State
  \Function{UpdateTheta}{forward samples $\mathcal{S}_F$, reverse samples $\mathcal{S}_R$} 
    \State Initialize $\mathcal{S}_{\theta,\mathrm{mb}} \gets \{ \}$
    \RepeatN{$N_\mathrm{mb}$} \hfill \Comment{Use larger $N_\mathrm{mb}$ for larger $|\mathcal{S}_F|$} 
      \State Randomly select $\mathcal{S}^{\mathrm{mb}}_F \subset \mathcal{S}_F$ of size $n^\mathrm{mb}_s$ without replacement
      \State Randomly select $\mathcal{S}^{\mathrm{mb}}_R \subset \mathcal{S}_R$ of size $n^\mathrm{mb}_s$ without replacement
      \State $\theta^* \gets \mathrm{argmin}_\theta \{ \hat{J}_F(\theta; \mathcal{S}^{\mathrm{mb}}_F ) +  \hat{J}_R(\theta; \mathcal{S}^{\mathrm{mb}}_R) \, | \, n_F^\mathrm{eff}(\theta; \mathcal{S}^{\mathrm{mb}}_F ) \geq f n^\mathrm{mb}_s, n_R^\mathrm{eff}(\theta; \mathcal{S}^{\mathrm{mb}}_R ) \geq f n^\mathrm{mb}_s \}$ 
      \State $\mathcal{S}_{\theta,\mathrm{mb}}  $.insert($\theta^*$)
    \End
    \State \Return mean($\mathcal{S}_{\theta,\mathrm{mb}} $)
  \EndFunction

  \State

  \Procedure{Main}\,
    \State Initialize parameters $\theta \gets \theta_\mathrm{init}$ and sample arrays $\mathcal{S}_F, \mathcal{S}_R \gets \{ \} , \{ \} $
    \Repeat
      \RepeatN{$N_\mathrm{s}$} \hfill \Comment{Use $N_\mathrm{mb} + N_\mathrm{s}$ on first iteration}
        \State $\mathcal{S}_F$.insert(\Call{RunTrajF}{$\theta$})
        \State $\mathcal{S}_R$.insert(\Call{RunTrajR}{$\theta$})
      \End
      \State Update estimate $\widehat{\Delta F} \gets \widehat{\Delta F}_\mathrm{BAR}(\mathcal{S}_F, \mathcal{S}_R)$ 
      \State $\theta \gets$ \Call{UpdateTheta}{$\mathcal{S}_F, \mathcal{S}_R$} 
    \Until out of computer time
  \EndProcedure
  
  \end{algorithmic}
\end{algorithm}
\end{figure}

\begin{figure}
\centering
\includegraphics[width=\textwidth]{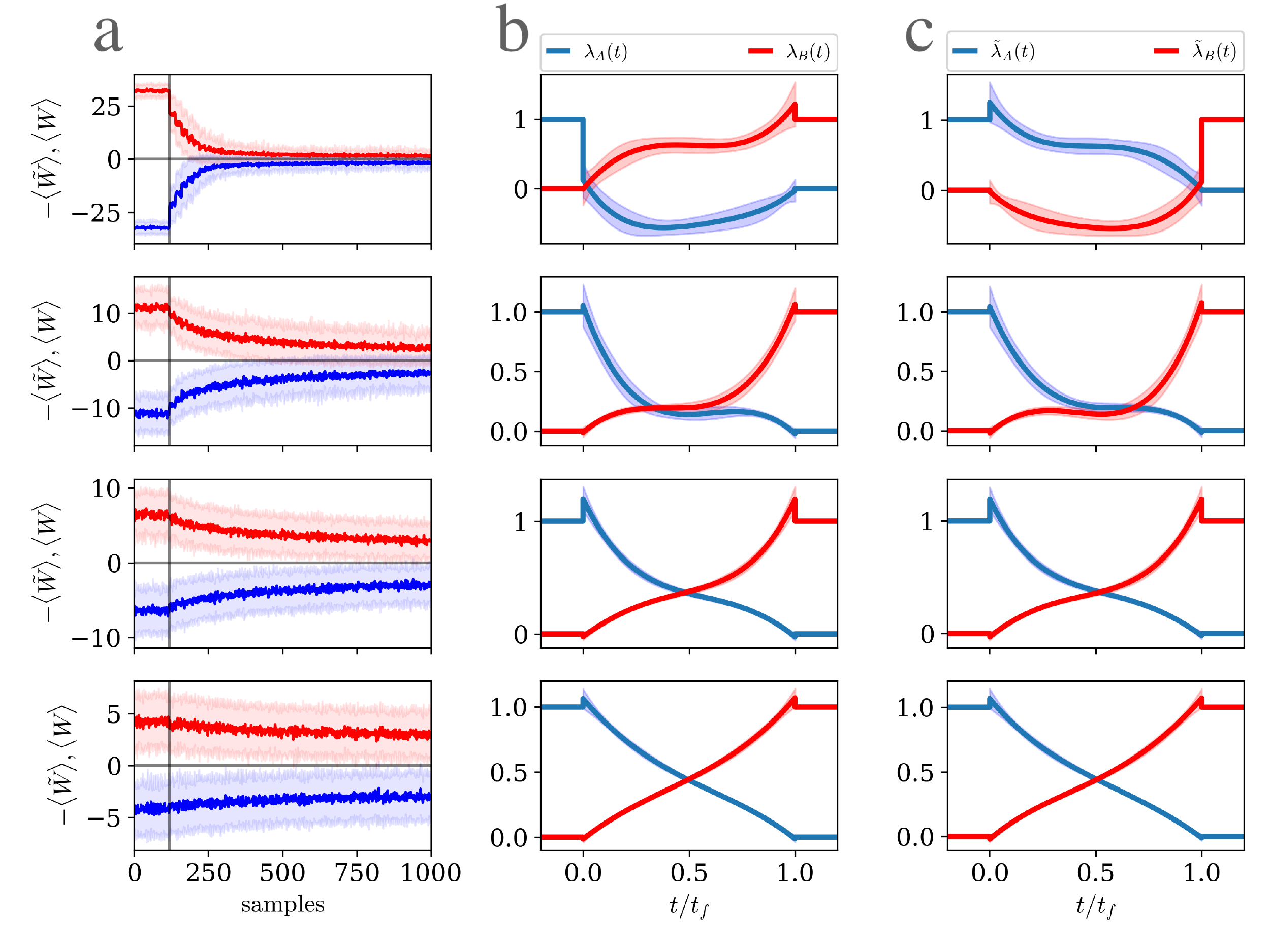}
    \caption{(a) For the linearly-biased double well, ensemble-averaged work across 100 trials, as a function of sample number. Parameter optimization begins at 120 samples, and happens every 20 samples. The rows correspond to protocol times $t_f = 0.2, 5.0, 20.0,$ and $50.0$ respectively. Convergence is slower for larger $t_f$; for $t_f = 50.0$ the protocol may not have converged within 1000 samples. (b) The forward protocols $U_F(\cdot, t) = \lambda_A(t) U_A(\cdot) + \lambda_B(t) U_B(\cdot)$ after 1000 samples. Rows correspond to the same $t_f$. (c) The reverse protocols $U_R(\cdot, t) = \tilde{\lambda}_A(t) U_A(\cdot) + \tilde{\lambda}_B(t) U_B(\cdot)$ after 1000 samples. The reverse protocols appear to satisfy $\tilde{\lambda}_A(t) = \lambda_B(t_f - t)$ and $\tilde{\lambda}_B(t) = \lambda_A(t)$, which is due to the symmetry of $U_A(\cdot)$ and $U_B(\cdot)$ in the problem.  }
    \label{fig:SI-DW}
\end{figure}
 
\begin{figure}
\centering
\includegraphics[width=\textwidth]{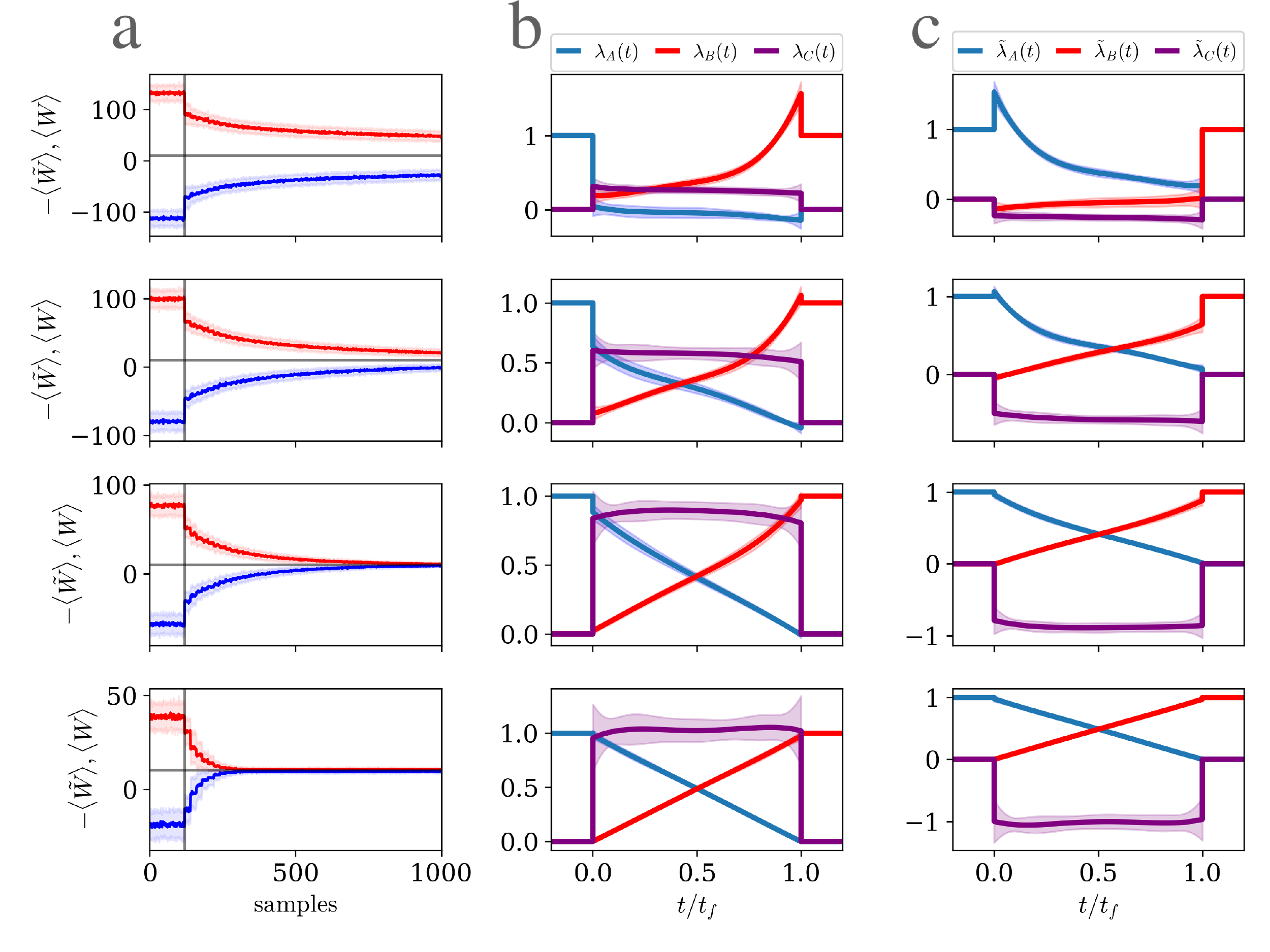}
    \caption{(a) For the Rouse polymer, ensemble-averaged work across 100 trials, as a function of sample number. Parameter optimization begins at 120 samples, and happens every 20 samples. The rows correspond to protocol times $t_f = 0.05 \, \tau_\mathrm{R} , 0.12 \, \tau_\mathrm{R} , 0.25 \, \tau_\mathrm{R},$ and $ 1.23 \, \tau_\mathrm{R}$ respectively. Unlike for the double well, convergence here is slower for smaller $t_f$. (b) The forward protocols $U_F(\cdot, t) = \lambda_A(t) U_A(\cdot) + \lambda_B(t) U_B(\cdot) + \lambda_C(t) U_C(\cdot) $ after 1000 samples. Rows correspond to the same $t_f$. For $t_f = 0.05 \, \tau_\mathrm{R}$ and $t_f = 0.12 \, \tau_\mathrm{R}$, the protocol has not yet converged to the counterdiabatic solution $\lambda_A(t) = (1 - t / t_f), \lambda_B(t) = t / t_f, \lambda_C = 1$. (c) The reverse protocols $U_R(\cdot, t) = \tilde{\lambda}_A(t) U_A(\cdot) + \tilde{\lambda}_B(t) U_B(\cdot) + \tilde{\lambda}_C(t) U_C(\cdot)$ after 1000 samples. The reverse protocols appear to satisfy $\tilde{\lambda}_A(t) = \lambda_B(t_f - t)$, $\tilde{\lambda}_B(t) = \lambda_A(t)$, and $\tilde{\lambda}_C(t) = -\lambda_C(t)$, which is due to the symmetry of $U_A(\cdot), U_B(\cdot)$, and $U_C(\cdot)$ in the problem. }
    \label{fig:SI-Rouse}
\end{figure}
 
\begin{figure}
\centering
\includegraphics[width=\textwidth]{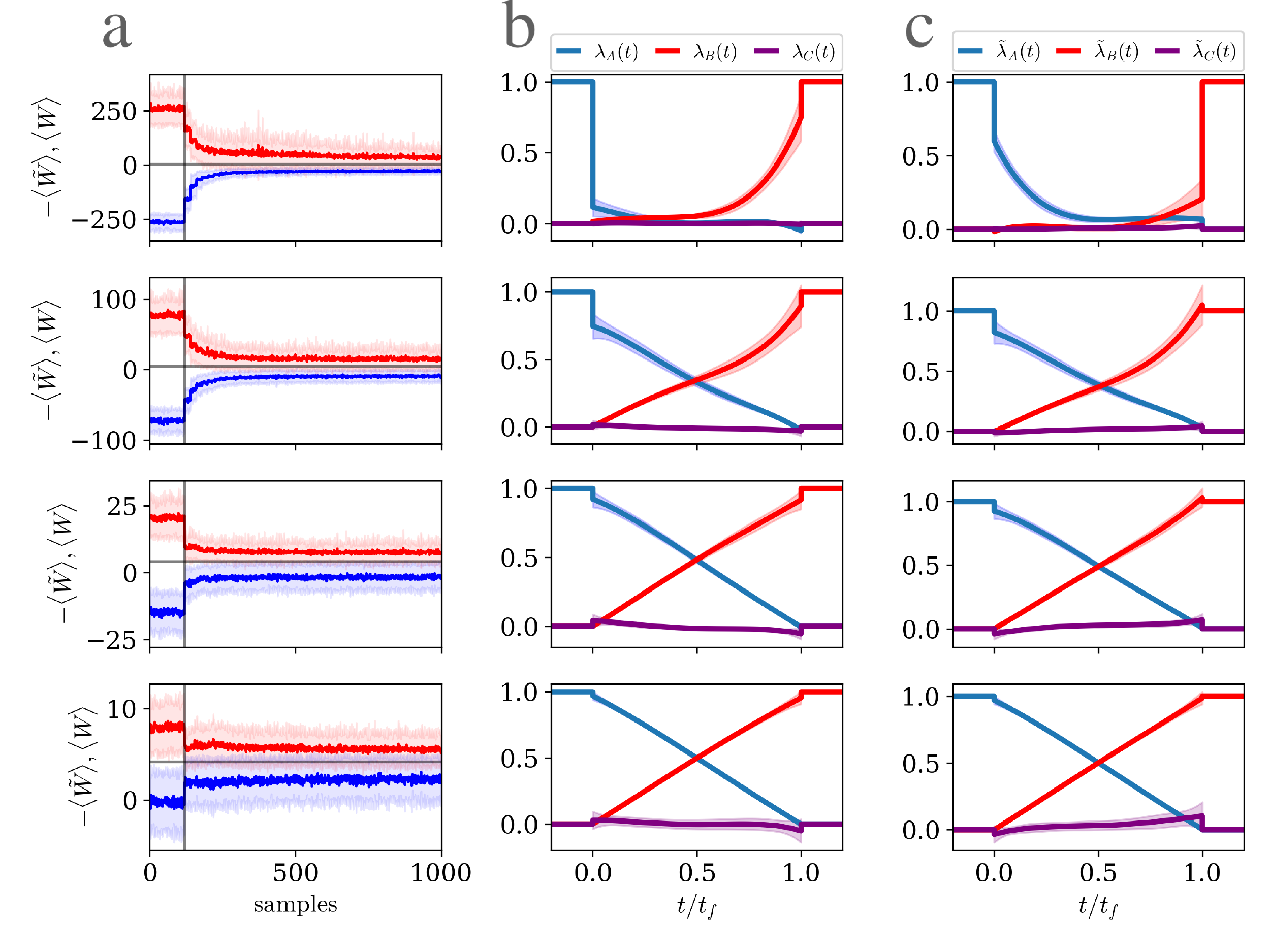}
    \caption{(a) For the worm-like chain, ensemble-averaged work across 100 trials, as a function of sample number. Parameter optimization begins at 120 samples, and happens every 20 samples. The rows correspond to protocol times $t_f = 0.07 \, \tau_\mathrm{LJ} , 0.28 \, \tau_\mathrm{LJ} , 1.41 \, \tau_\mathrm{LJ},$ and $7.07 \, \tau_\mathrm{LJ}$ respectively. It appears convergence is reached rapidly, within the 1000 samples for all cases. (b) The forward protocols $U_F(\cdot, t) = \lambda_A(t) U_A(\cdot) + \lambda_B(t) U_B(\cdot) + \lambda_C(t) U_C(\cdot) $ after 1000 samples. Rows correspond to the same $t_f$. For small $t_f$, the protocol has reduced magnitude. This corresponds to lowering the potential, or raising the temperature (i.e., smaller $\beta U(\cdot, t)$). (c) The reverse protocols $U_R(\cdot, t) = \tilde{\lambda}_A(t) U_A(\cdot) + \tilde{\lambda}_B(t) U_B(\cdot) + \tilde{\lambda}_C(t) U_C(\cdot)$ after 1000 samples. Due to the intrinsic asymmetry of the problem between pulled and collapsed states, the resulting reverse protocols do not obey the symmetries observed in the double-well and Rouse polymer problems.}
    \label{fig:SI-WLC}
\end{figure}
 
\begin{figure}
\centering
\includegraphics[width=\textwidth]{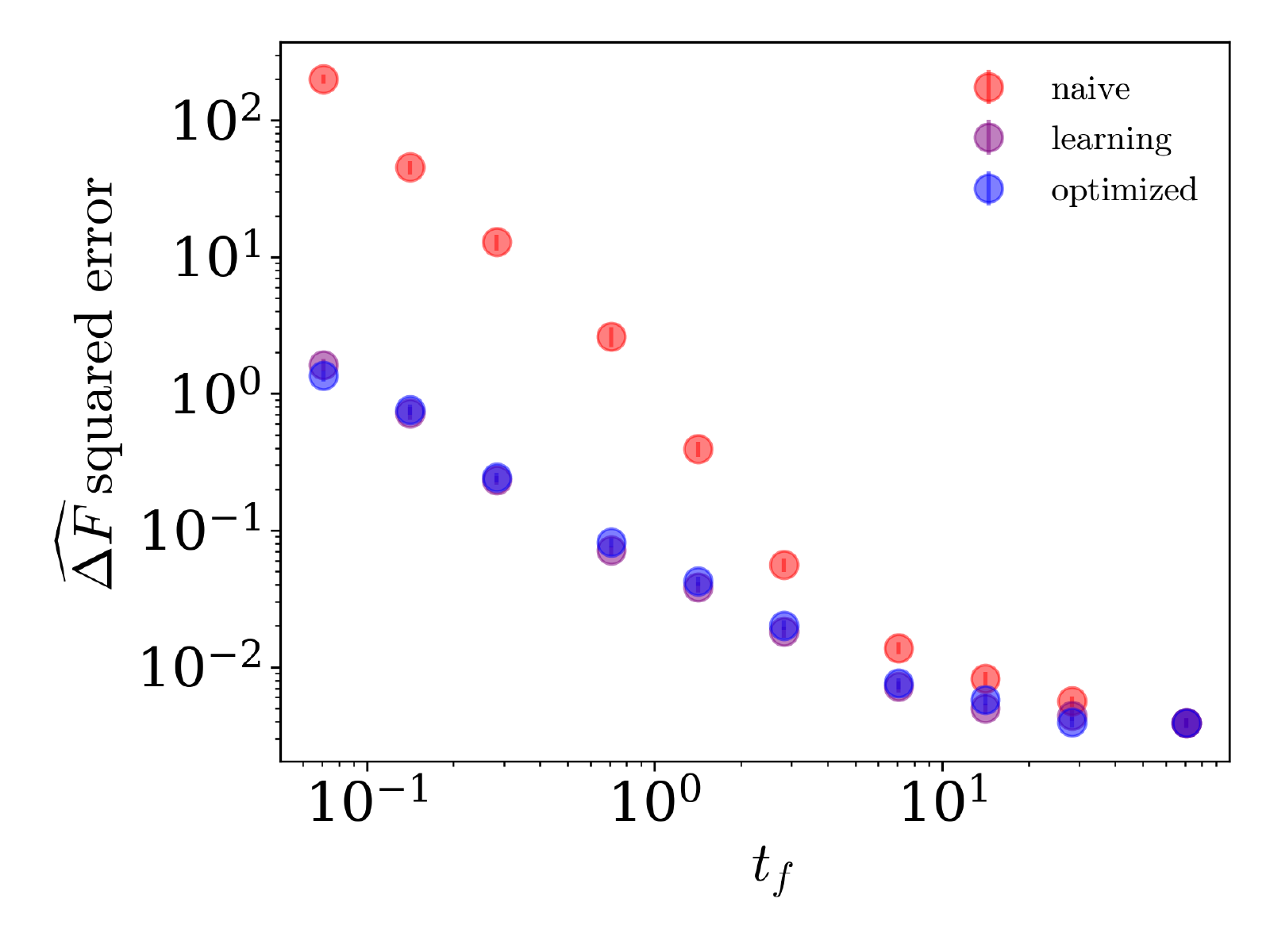}
    \caption{Performance plot for the worm-like chain. That the mean squared error for protocol learning is near equal to the optimized protocol implies convergence occurs quickly within protocol optimization, cf Fig. \ref{fig:SI-WLC}(a). At $t_f = 0.07\, \tau_\mathrm{LJ}$, the MSE is $123.3$ times lower under protocol optimization than under the naive protocol.  }
    \label{fig:SI-WLC-performance}
\end{figure}

\bibliography{main}